\documentclass[floats,superscriptaddress,twocolumn,pra]{revtex4-2}
\usepackage{graphics}
\usepackage{amsmath}
\usepackage{graphicx}
\usepackage{amsfonts}
\usepackage{amssymb}
\usepackage{eurosym}
\usepackage{graphics}
\usepackage{float}
\usepackage{longtable}
\usepackage{epsfig}
\usepackage{latexsym}
\usepackage{theorem}
\usepackage{bm}
\usepackage{ulem}
\usepackage{color}
\newcommand{\SP}[1]{{{\textcolor{black}{#1}}}}

\begin{document}

\title{Improved composable key rates for CV-QKD}
\author{Stefano Pirandola}
\affiliation{Computer Science, University of York, York YO10 5GH
(UK)}
\author{Panagiotis Papanastasiou}
\affiliation{nodeQ, The Catalyst, Baird Lane, York, YO10 5GA
(UK)}

\begin{abstract}
Modern security proofs of quantum key distribution (QKD) must take finite-size effects and composable aspects into consideration. This is also the case for continuous-variable (CV) protocols which are based on the transmission and detection of bosonic coherent states. In this paper, we refine and advance the previous theory in this area providing a more rigorous formulation for the composable key rate of a generic CV-QKD protocol. Thanks to these theoretical refinements, our general formulas allow us to prove more optimistic key rates with respect to previous literature.
\end{abstract}
\maketitle

\section{Introduction}
Quantum key distribution (QKD) is arguably one of the most advanced areas in quantum information,
both theoretically and experimentally~\cite{AOP,Feihu,Gisin}, with very well-known limits, such as the fundamental PLOB bound for repeater-less quantum communication~\cite{PRL2009,PLOB} and its extension to repeaters and networks with arbitrary topologies and routing mechanisms~\cite{ETEbounds}. In particular, the continuous-variable (CV) version of QKD is a preferred option that has been gradually improved in various aspects, such as the rigor of the security proofs,
the speed of data processing techniques, and the distance of experimental implementations~\cite[Secs.~7 and~8]{AOP}.
In terms of CV-QKD theory, the first asymptotic analyses were extended to finite-size effects and, later, to composable security proofs~\cite{Ralph99,Hillery00,Cerf01,Ralph02,Fred02,Fred06,Pirandola08, Lev2010, Vlad14, Lev2015, Vlad15, Vlad16, Lev2017, Pirandola21, Pirandola21b, Panos21, Matsuura2021,Alex_I,Alex_II,MasoudI,MasoudII} (see also Ref.~\cite[Sec.~9]{AOP}).

Here we build on previous composable security analyses of CV-QKD~\cite{Pirandola21,Pirandola21b,Panos21} to provide a more refined and advanced formulation. Our revised formulas enables us to achieve more optimistic key rates for CV-QKD than previous literature. The results apply to a variety of protocols, including schemes with discrete-alphabet or continuous (Gaussian~\cite{Fred03, Weed04}) modulation of coherent states, with homodyne or heterodyne detection, CV measurement device independent (MDI) QKD~\cite{CVMDI,Alex_III}, and also the post-selection versions of these protocols.

The paper is structured as follows. In Sec.~\ref{SEC2} we derive our general formula for the secret key rate of a generic CV-QKD protocol; this is done by refining previous theory and adopting a number of improvements, including a different approach to tensor-product reduction after error correction (proven in Appendix~\ref{appEXT}). In Sec.~\ref{SEC3}, we apply the results to relevant examples of CV-QKD protocols, showing the improvements in terms of key rate with respect to previous literature. Sec.~\ref{sec:clarifications} contains some clarifications and Sec.~\ref{SEC4} is for conclusions.

\section{Composable key rate\label{SEC2}}
In this section we derive an improved formula for the secret key rate of a
generic CV-QKD protocol in the finite-size and composable
framework. The main derivation is performed under the assumption
of collective attacks, but the result will be easily extended to
coherent attacks in the case of one of the Gaussian-modulated
protocols. We present the various ingredients and aspects of the proof 
in a number of subsections.

\subsection{Output state of a CV-QKD protocol}
Consider a CV-QKD protocol where $N$ single-mode systems are
transmitted from Alice $A$ to Bob $B$. A portion $n$ of these
systems will be used for key generation, while a portion $m=N-n$
will be used for parameter estimation. Let us assume that the
bosonic communication channel depends on a number $n_{\text{pm}}$
of parameters $\mathbf{p}=(p_1,p_2,\ldots)$ (e.g., transmissivity
and thermal noise of the channel). These parameters are estimated
by the parties and we will account for their partial knowledge at
the end of the derivations. For now, let us assume that Alice and
Bob has perfect knowledge of $\mathbf{p}$.

Under the action of a collective attack, the output classical-quantum (CQ) state of
Alice ($A$), Bob ($B$) and Eve ($E$) has the tensor-structure form
$\rho^{\otimes n}$, where%
\begin{equation}
\rho=\sum_{k,l \in \{0,\dots, 2^{d}-1\}} p(k,l)\left\vert k\right\rangle _{A}\left\langle k\right\vert \otimes
\left\vert l\right\rangle _{B}\left\langle l\right\vert \otimes\rho_{E}^{k,l}.
\label{sharedSSS}
\end{equation}
Here Alice's variable $k$ and Bob's variable $l \in \mathcal{L}=\{0,\dots,2^d-1\}$ are both multi-ary symbols ($2^d$-ary, equivalent to $d$-bit strings)
and $p(k,l)$ is their joint probability distribution (depending on the interaction used by Eve).

In the case of a protocol based on the Gaussian modulation of coherent states, the multi-ary symbols are the output of analog-to-digital conversion (ADC) from Alice's and Bob's quadratures, $x$ and $y$, i.e., we have $x\overset{\text{ADC}}{\rightarrow}k$ and $y\overset{\text{ADC}}{\rightarrow}l$. If the protocol is based on the homodyne detection, we have that $y$ is randomly created by a random switching between the $q$ and $p$ quadrature (with Alice choosing the corresponding quadrature for each instance, upon Bob's classical communication). If the protocol is based on the heterodyne detection, both $q$ and $p$ quadratures are used, so we have
$y = (q_{B},p_{B})\overset{\text{ADC}}{\rightarrow}(l_{q},l_{p})$ followed by the concatenation $l=l_{q}l_{p}$ so that the dimension is $d=d_q+d_p$, where $d_q$ ($d_p$) refers to the dimension of $l_q$ ($l_p$). 
%** HERE TO CLARIFY ABOUT d for Heterodyne **
%(this is also the approach for CV-MDI).
Finally, in the case of CV-QKD protocols based on discrete-alphabet coherent states, no ADC is necessary and the discretized variables are directly expressed by the encoding variables.

Whatever protocol is used, after $n$ uses, there will be two
sequences of multi-ary symbols, $\mathbf{k}=(k_1,k_2,\ldots)$ and
$\mathbf{l}=(l_1,l_2,\ldots)$, each with length $n$ (so their
 equivalent binary length would be $n d$). These are generated with
joint probability $p(\mathbf{k},\mathbf{l}) = \prod_{i=1}^{n}
p(k_i,l_i)$, and the total $n$-use state of Alice, Bob and Eve
reads
\begin{equation}
\rho^{\otimes n}=\sum_{\mathbf{k},\mathbf{l}~\in \{0,\dots,2^d-1\}^n} p(\mathbf{k},\mathbf{l}) \left\vert \mathbf{k} \right\rangle _{A^n}\left\langle \mathbf{k} \right\vert \otimes \left\vert \mathbf{l} \right\rangle _{B^n}\left\langle \mathbf{l} \right\vert \otimes\rho_{E^n}^{\mathbf{k},\mathbf{l}},
\label{sharedSSS2}
\end{equation}
where $\left\vert \mathbf{k} \right\rangle = \otimes_{i=1}^{n} \left\vert k_i \right\rangle$,  $\left\vert \mathbf{l} \right\rangle = \otimes_{i=1}^{n} \left\vert l_i\right\rangle$ and
\begin{equation}
%\left\vert k^n (l^n)\right\rangle = \otimes_{i=1}^{n} \left\vert k_i (l_i)\right\rangle,~
%\left\vert l^n\right\rangle = \otimes_{i=1}^{n} \left\vert l_i\right\rangle,~
\rho_{E^n}^{\mathbf{k},\mathbf{l}} = \otimes_{i=1}^{n} \rho_{E}^{k_i,l_i}.
\end{equation}
\subsection{Error correction and epsilon correctness \label{EC}}

Alice and Bob will then perform procedures of error correction
(EC) and privacy amplification (PA) over the state $\rho^{\otimes
n}$ with the final goal to approximate the $s_{n}$-bit ideal CQ
state, which is of the type
\begin{align}\label{rho_ideal}
& \rho^{n}_{\text{ideal}}= \omega^{n}_{AB} \otimes \rho_{E^n}, \\
&\omega^{n}_{AB} :=2^{-s_{n}} \sum_{\mathbf{s}} \left\vert
\mathbf{s} \right\rangle _{A^{n}}\left\langle \mathbf{s}
\right\vert \otimes\left\vert \mathbf{s} \right\rangle
_{B^{n}}\left\langle \mathbf{s} \right\vert,
\end{align}
where Alice's and Bob's classical systems contain the same random
binary sequence $\mathbf{s}$ of length $s_{n}$, from which Eve is
completely decoupled (note that the final output is a binary
sequence even if we start from multi-ary symbols $k$ and $l$ for
Alice and Bob).

%%%%%%%%%%%%%%%%%%%%%%%%%%%%%%%%%%%%%%%%%%%%%%%%%%%%%%%%%%%%%%%%%

In reverse reconciliation, Alice attempts to reconstruct Bob's
sequence $\mathbf{l}$. During EC, Bob publicly reveals
$\mathrm{leak}_{\text{ec}}$ bits of information to help Alice to
compute her guess $\hat{\mathbf{l}}$ of $\mathbf{l}$ starting from
her local data $\mathbf{k}$. In practical schemes of EC (based on
linear codes, such as LDPC codes), these
$\mathrm{leak}_{\text{ec}}$ bits of information corresponds to a
syndrome $\mathrm{synd}(\bold{l})$ that Bob computes over his sequence $\mathbf{l}$,
interpreted as a noisy codeword of a linear code agreed with
Alice.

Then, as a verification, Alice and Bob publicly compare hashes
computed over $\mathbf{l}$ and $\hat{\mathbf{l}}$. If these hashes
coincide, the two parties go ahead with the probability
$p_{\text{ec}}$, otherwise, they abort the protocol. We denote by
$T_\text{ec}$ the case of a successful verification (no abort), so
that $\rho_{|T_\text{ec}}$ represents a conditional post-EC state.
More specifically, the hash comparison requires Bob to send
$\left\lceil -\log_{2}\varepsilon _{\text{cor}}\right\rceil $ bits
to Alice for some suitable $\varepsilon _{\text{cor}}$ (the number
of these bits is typically small in comparison to
$\mathrm{leak}_{\text{ec}}$). Parameter $\varepsilon_{\text{cor}}$
is called the $\varepsilon$-correctness~\cite[Sec.~4.3]{Portmann}
and it bounds the probability that Alice's and Bob's corrected
sequences are different. The probability of such an error is bounded by~\cite{TomaEpsCorr}
\begin{equation}
p_{\text{ec}}\mathrm{Prob}(\hat{\mathbf{l}} \neq \mathbf{l} \vert T_\text{ec})
\leq\varepsilon_{\text{cor}}. \label{ecorrectness}
\end{equation}
% \begin{equation}
% p_{\text{ec}}\mathrm{Prob}(\hat{\mathbf{l}} \neq \mathbf{l} \vert T_\text{ec})\leq p_{\text{ec}
% }2^{-\left\lceil -\log_{2}\varepsilon_{\text{cor}}\right\rceil}
% \leq\varepsilon_{\text{cor}}. \label{ecorrectness}
% \end{equation}

\subsection{Equivalence to a projection process\label{equiv_proj}}
As discussed above, EC consists of two steps. In the first (correction) step, Bob sends the syndrome information $\text{synd}(\mathbf{l})$ to Alice. Conditionally on $\text{synd}(\mathbf{l})$, she transforms her
variable via a function 
\begin{equation}
  \mathbf{k} \mapsto
f_\text{guess}(\mathcal{\mathbf{k}},\text{synd}(\mathbf{l}))=\hat{\mathbf{l}} \in \{0,2^d-1\}^n. \label{transformA}
\end{equation}
%This can also be represented by a (conditional) classical channel $\mathcal{E}_\text{guess}(|\mathbf{k}\rangle\langle \mathbf{k}|)= |\hat{\mathbf{l}}\rangle\langle \hat{\mathbf{l}}|$.
The second (verification) step is the verification of
the hashes. If successful, this is equivalent
%to Alice and Bob identifying original sequences $(\mathbf{k},\mathbf{l})\in \Gamma$ such that
to having a corrected sequence $\hat{\mathbf{l}}$
that is indistinguishable from $\mathbf{l}$ with a probability
larger than $1-\varepsilon_\text{cor}$. 

Overall, successful EC is equivalent to filtering the entire set of initial
sequences $(\mathbf{k},\mathbf{l})~\in\{0,\dots,2^d-1\}^n \otimes \{0,\dots,2^d-1\}^n $ into a
subset of ``good'' sequences
% \begin{equation}
% \Gamma=\{(\mathbf{k},\mathbf{l}): \mathrm{Prob}(\hat{\mathbf{l}}\neq
% \mathbf{l} \vert T_\text{ec})\leq \varepsilon_\text{cor}\},
% \end{equation}
\begin{equation}
\Gamma=\{(\mathbf{k},\mathbf{l}): \mathrm{Prob}(\mathbf{k}\neq
\mathbf{l} )\leq \varepsilon_\text{cor}\},
\end{equation}
with associated probability $p_\text{ec}=\sum_{(\mathbf{k},\mathbf{l})\in \Gamma}p(\mathbf{k},\mathbf{l})$. This can equivalently be represented by a projection
\begin{equation}\label{EC-proj}
\rho^{\otimes n} \rightarrow \Pi_{\Gamma} \rho^{\otimes n}
\Pi_{\Gamma},~~\Pi_{\Gamma} = \sum_{(\mathbf{k},\mathbf{l}) \in
\Gamma}\left\vert \mathbf{k}, \mathbf{l} \right\rangle
\left\langle \mathbf{k}, \mathbf{l} \right\vert,
\end{equation}
restricting the classical states to the labels
$(\mathbf{k},\mathbf{l}) \in \Gamma$ followed by the application of the quantum operation
\begin{equation}
    \cal{E}_{\mathrm{guess}}(\left\vert \mathbf{k}, \mathbf{l} \right\rangle
\left\langle \mathbf{k}, \mathbf{l} \right\vert) = \vert  \hat{\mathbf{l}}, \mathbf{l} \rangle
\langle \hat{\mathbf{l}} , \mathbf{l} \vert, \label{chanMain}
\end{equation}
according to the transformation in Eq.~\eqref{transformA}. In particular, note that this operation is a completely positive trace-preserving (CPTP) map, i.e., a quantum channel.
%of $\mathcal{E}_\text{guess}$, i.e., $\mathcal{E}_\text{guess}(\Pi_{\Gamma} \rho^{\otimes n} \Pi_{\Gamma})$. 

Thus, the (normalized) post-EC state is given by
\begin{align}\label{EC_state}
\tilde{\rho}_{ABE|T_\text{ec}}^{n}=&\sum_{\substack{(\mathbf{k},\mathbf{l})\in \Gamma\\ \hat{\mathbf{l}}=f_\text{guess}(\mathbf{k},\text{synd}(\mathbf{l}))}} \frac{p(\mathbf{k},\mathbf{l})}{p_\text{ec}} \vert \hat{\mathbf{l}},\mathbf{l} \rangle _{A^nB^n}\langle \hat{\mathbf{l}},\mathbf{l}\vert  \otimes\rho_{E^n}^{\mathbf{k},\mathbf{l}}.
\end{align}
It is clear that the state above, expressed in terms of $n$-long sequences of $2^d$-ary symbols, can equivalently be rewritten in terms of $nd$-long binary strings. It is also important to note that, due to the projection, the state after EC no longer has a tensor product structure.

\subsection{Privacy amplification and epsilon secrecy}
The next step is PA which realizes the randomness extraction while
decoupling Eve. The parties agree to use a function $f$ randomly
chosen from a family $F$ of 2-universal hash functions with
probability $p(f)$ among a total of $|F|$ possible choices (note
that it is necessary to randomize over the hash functions as
discussed in Ref.~\cite{Bennett95}).
Then, they transform their multi-ary $n$-long sequences into $nd$-long binary strings (so the state in Eq.~\eqref{EC_state} is suitably expressed in terms of these binary strings). Such strings are individually compressed into a key pair \{$\hat{\mathbf{s}},\mathbf{s}\}$ of $s_n < nd$ random bits. %In this way they compress their sequences \{$\hat{\mathbf{l}},\mathbf{l}\}$ into a key pair \{$\hat{\mathbf{s}},\mathbf{s}\}$ of $s_n < n$ random bits.

The process of PA can be described by a CPTP map $\rho_F \otimes
\tilde{\rho}_{ABE|T_\text{ec}}^{ n} \rightarrow
\bar{\rho}^{n}_{ABEF|T_\text{ec}}$, where
\begin{align}
  &\bar{\rho}^{n}_{ABEF|T_\text{ec}} =  p_{\mathrm{ec}}^{-1}\sum_{f,\hat{\mathbf{s}},\mathbf{s}} p(f) p(\hat{\mathbf{s}},\mathbf{s}) \times  \nonumber  \\
  & \vert \hat{\mathbf{s}}\rangle _{A^n}\langle \hat{\mathbf{s}}\vert  \otimes
\vert \mathbf{s}\rangle _{B^n}\langle \mathbf{s}\vert
\otimes\rho_{E^n}^{f,\hat{\mathbf{s}},\mathbf{s}} \otimes \vert
f\rangle _{F}\langle f\vert,
\end{align}
which is a generalization of Ref.~\cite[Eq.~(5.5)]{RennerThesis}.
This also means that Alice's and Bob's sequences undergo local
data processing which cannot increase their distinguishability,
i.e., we have
\begin{equation}
   \mathrm{Prob}(\hat{\mathbf{s}} \neq \mathbf{s}|T_\text{ec}) \leq \mathrm{Prob}(\hat{\mathbf{l}} \neq \mathbf{l} |T_\text{ec}),\label{prob_connection}
\end{equation}
due to the pigeonhole principle. By tracing out Alice, we can write the reduced state of Bob (containing the key) and Eve
\begin{equation}
\bar{\rho}^{n}_{BEF|T_\text{ec}} =
p_{\mathrm{ec}}^{-1}\sum_{f,\hat{\mathbf{s}},\mathbf{s}} p(f)
p(\hat{\mathbf{s}},\mathbf{s}) \vert \mathbf{s}\rangle
_{B^n}\langle \mathbf{s}\vert
\otimes\rho_{E^n}^{f,\hat{\mathbf{s}},\mathbf{s}} \otimes \vert
f\rangle _{F}\langle f\vert.
\end{equation}
On the latter state, we impose the condition of
$\varepsilon$-secrecy for Bob. First note that we may write the
ideal state as $\omega^{n}_{B} \otimes \bar{\rho}^{n}_{EF|T_\text{ec}}$,
where
\begin{equation}
 \omega^{n}_{B}
:=2^{-s_{n}} \sum_{\mathbf{s}}\left\vert \mathbf{s}
\right\rangle_{B^{n}}\left\langle \mathbf{s}
\right\vert,~\bar{\rho}^{n}_{EF|T_\text{ec}}:=
\mathrm{Tr}_{B}(\bar{\rho}^{n}_{BEF|T_\text{ec}}).
\end{equation}
Then, we impose that the distance from this ideal state must be
less than $\varepsilon_{\text{sec}}$, i.e., we impose
\begin{equation}
   p_{\text{ec}}D(\bar{\rho}^{n}_{BEF|T_\text{ec}},\omega^{n}_{B} \otimes \bar{\rho}^{n}_{EF|T_\text{ec}})\leq\varepsilon_{\text{sec}}. \label{esecrecy}
\end{equation}
% \bar{\rho}^{n}_{EF|T_\text{ec}}
% \omega^{n}_{AB} \otimes \bar{\rho}^{n}_{EF|T_\text{ec}}

\subsection{Combining correctness and secrecy into epsilon security}
Following Ref.~\cite[Th.~4.1]{Portmann}, we can combine the features of correctness and secrecy into a single epsilon parameter.
In fact, if Eqs.~\eqref{ecorrectness} and~\eqref{esecrecy} hold, then we may write the condition for $\varepsilon$-security for Alice and Bob
\begin{equation}
p_{\text{ec}}D(\bar{\rho}^{n}_{ABEF|T_\text{ec}},\omega^{n}_{AB}
\otimes \bar{\rho}^{n}_{EF|T_\text{ec}})\leq
\varepsilon:=\varepsilon_{\text{cor}}+\varepsilon_{\text{sec}}.
\label{esecurity}
\end{equation}
It is instructive to repeat the proof of this result from
Ref.~\cite[Sec.~4.3]{Portmann}.

\bigskip
\textbf{Proof.}~~Let us define the following state, similar to
$\bar{\rho}^{n}_{ABEF|T_\text{ec}}$ but where Alice's system is
copied from Bob's so they have exactly the same key string
\begin{align}
  &\bar{\gamma}^{n}_{ABEF|T_\text{ec}} =  p_{\mathrm{ec}}^{-1}\sum_{f,\hat{\mathbf{s}},\mathbf{s}} p(f) p(\hat{\mathbf{s}},\mathbf{s})   \times \nonumber  \\
  & \vert \mathbf{s}\rangle _{A^n}\langle \mathbf{s}\vert  \otimes
\vert \mathbf{s}\rangle _{B^n}\langle \mathbf{s}\vert \otimes\rho_{E^n}^{f,\hat{\mathbf{s}},\mathbf{s}} \otimes \vert f\rangle _{F}\langle f\vert.
\end{align}
Then, we can use the triangle inequality to write
\begin{align}\label{triangle_ineq}
    &D(\bar{\rho}^{n}_{ABEF|T_\text{ec}},\omega^{n}_{AB} \otimes \bar{\rho}^{n}_{EF|T_\text{ec}}) \leq \nonumber \\
    &D(\bar{\rho}^{n}_{ABEF|T_\text{ec}},\bar{\gamma}^{n}_{ABEF|T_\text{ec}})\nonumber
    \\
    &+D(\bar{\gamma}^{n}_{ABEF|T_\text{ec}},\omega^{n}_{AB} \otimes \bar{\rho}^{n}_{EF|T_\text{ec}}).
\end{align}
The first term accounts for the correctness and can be bounded as follows
\begin{align}\label{correctness}
&D(\bar{\rho}^{n}_{ABEF|T_\text{ec}},\bar{\gamma}^{n}_{ABEF|T_\text{ec}})  \nonumber \\
&\leq  p_{\mathrm{ec}}^{-1}\sum_{f,\hat{\mathbf{s}},\mathbf{s}} p(f) p(\hat{\mathbf{s}},\mathbf{s}) D(\vert \hat{\mathbf{s}}\rangle _{A^n}\langle \hat{\mathbf{s}}\vert, \vert \mathbf{s}\rangle _{A^n}\langle \mathbf{s}\vert)  \nonumber \\
&= \sum_{\hat{\mathbf{s}} \neq \mathbf{s}} \frac{p(\hat{\mathbf{s}},\mathbf{s})}{p_{\mathrm{ec}}} \nonumber \\
&= \mathrm{Prob}(\hat{\mathbf{s}} \neq \mathbf{s} |T_\text{ec})  \nonumber \\
&\leq \mathrm{Prob}(\hat{\mathbf{l}} \neq \mathbf{l}|T_\text{ec}).
\end{align}
The second term in Eq.~(\ref{triangle_ineq}) accounts for secrecy
and can be manipulated as follows
\begin{align}
 &D(\bar{\gamma}^{n}_{ABEF|T_\text{ec}},\omega^{n}_{AB} \otimes \bar{\rho}^{n}_{EF|T_\text{ec}}) \nonumber \\
 &=D(\bar{\gamma}^{n}_{BEF|T_\text{ec}},\omega^{n}_{B} \otimes \bar{\rho}^{n}_{EF|T_\text{ec}}) \nonumber \\
 &=D(\bar{\rho}^{n}_{BEF|T_\text{ec}},\omega^{n}_{B} \otimes
 \bar{\rho}^{n}_{EF|T_\text{ec}}),
\end{align}
where we use the fact that the trace distance does not change if
we trace Alice's cloned system in $\bar{\gamma}^{n}$.

Thus we have
\begin{align}
&p_{\text{ec}}D(\bar{\rho}^{n}_{ABEF|T_\text{ec}},\omega^{n}_{AB} \otimes \bar{\rho}^{n}_{EF|T_\text{ec}}) \nonumber \\
&\leq p_{\text{ec}} \mathrm{Prob}(\hat{\mathbf{l}} \neq \mathbf{l} |T_\text{ec}) \nonumber \\
&+ p_{\text{ec}} D(\bar{\rho}^{n}_{BEF|T_\text{ec}},\omega^{n}_{B} \otimes
\bar{\rho}^{n}_{EF|T_\text{ec}}).  \label{proven}
\end{align}
Using Eqs.~\eqref{ecorrectness} and~\eqref{esecrecy} in the right-hand side of Eq.~\eqref{proven} we get Eq.~\eqref{esecurity}.~$\blacksquare$

\subsection{Leftover hash bound}
We may now bound the distance of the privacy amplified state
$\bar{\rho}^{n}_{BEF|T_\text{ec}}$ from the ideal state
$\omega^{n}_{B} \otimes \bar{\rho}^{n}_{EF|T_\text{ec}}$
containing $s_n$ random and decoupled bits. For this, we employ
the converse leftover hash bound. % but in a form which is improved with respect to its use in Ref.~\cite{Pirandola21}.
Following Ref.~\cite[Th.~6]{TomaRenner}, we may write
\begin{align}
&p_{\text{ec}} D(\bar{\rho}^{n}_{BEF|T_\text{ec}},\omega^{n}_{B} \otimes \bar{\rho}^{n}_{EF|T_\text{ec}})\notag\\
&\leq \varepsilon_{\text{s}} +\frac{1}{2}
\sqrt{2^{s_n-H_{\text{min}}^{\varepsilon_{\text{s}}}(B^{n}|E^{n})_{\sigma^{
n}}}},\label{differ}
\end{align}
where $\sigma^{n}$ is Bob and Eve's sub-normalized state before PA
and after EC, given by
\begin{align}
\sigma^{n} & := \sigma_{BE|T_\text{ec}}^{n} =
 p_\text{ec}
\tilde{\rho}^{n}_{BE|T_\text{ec}} \nonumber
\\
& =\mathrm{Tr}_{A}[\mathcal{E}_\text{guess} (\Pi_{\Gamma}
\rho_{ABE}^{\otimes n}
\Pi_{\Gamma})] \nonumber
\\
& = \sum_{(\mathbf{k},\mathbf{l})\in \Gamma}
p(\mathbf{k},\mathbf{l}) \left\vert \mathbf{l} \right\rangle
_{B^n}\left\langle \mathbf{l} \right\vert
\otimes\rho_{E^n}^{\mathbf{k},\mathbf{l}}.\label{joint_state}
\end{align}
By imposing the condition
\begin{equation}
\varepsilon_{\text{s}} +\frac{1}{2}
\sqrt{2^{s_n-H_{\text{min}}^{\varepsilon_{\text{s}}}(B^{n}|E^{n})_{\sigma^{
n}}}} \leq \varepsilon_{\text{sec}}, \label{torearrange_main}
\end{equation}
we certainly realize the secrecy bound in Eq.~(\ref{esecrecy}). If
we also impose the condition of correctness in
Eq.~\eqref{ecorrectness}, we reach the condition of epsilon
security for Alice and Bob expressed by Eq.~\eqref{esecurity}.
Setting
$\varepsilon_{\text{h}}:=\varepsilon_{\text{sec}}-\varepsilon_{\text{s}}$
and re-arranging Eq.~\eqref{torearrange_main}, we derive the
following upper-bound for the binary length of the key (converse
leftover hash bound)
\begin{equation}
s_{n}\leq
H_{\text{min}}^{\varepsilon_{\text{s}}}(B^{n}|E^{n})_{\sigma^{n}}+2
\log_2(2\varepsilon_{\text{h}}).
\label{boundkey}%
\end{equation}
Thus, for the protocol to be epsilon-secure  with
$\varepsilon:=\varepsilon_{\text{cor}}+\varepsilon_{\text{sec}}=\varepsilon_{\text{cor}}+\varepsilon_{\text{s}}+\varepsilon_{\text{h}}$,
the binary length of the key cannot exceed the right-hand side of
Eq.~\eqref{boundkey}.

\subsection{Including the leakage due to EC \label{sec_leak}}
Let us better describe Eve's system $E^{n}$ as $E^{n}R$, where
$E^{n}$ are the systems used by Eve
during the quantum communication while $R$ is an extra register of dimension $\mathrm{dim}_{R}%
=2^{\mathrm{leak}_{\text{ec}}+\left\lceil -\log_{2}\varepsilon
_{\text{cor}}\right\rceil}$. The latter is used by Eve to store
the bits that are leaked during EC. This means that
Eq.~(\ref{boundkey}) is more precisely given by
\begin{equation}
s_{n}\leq
H_{\text{min}}^{\varepsilon_{\text{s}}}(B^{n}|E^{n}R)_{\sigma^{n}}+2
\log_2(2\varepsilon_{\text{h}}).
\label{boundkeyR}%
\end{equation}

We can then use Ref.~\cite[Prop.~5.10]{TomaThesis} for the smooth
min-entropy computed over generally sub-normalized states which
leads to
\begin{align}\label{Eve_classical_reg}
& H_{\text{min}}^{\varepsilon_{\text{s}}}(B^{n}|E^{n}
R)_{\sigma^{n}}\geq H_{\text{min}}^{\varepsilon_{\text{s}}}(B^{n} |E^{n})_{\sigma^{n}}-\log_{2}\mathrm{dim}_{R} \nonumber \\
&= H_{\text{min}}^{\varepsilon_{\text{s}}}(B^{n}
|E^{n})_{\sigma^{n}}-\mathrm{leak}_{\text{ec}}-\left\lceil
-\log_{2}\varepsilon
_{\text{cor}}\right\rceil \nonumber  \\
&\ge H_{\text{min}}^{\varepsilon_{\text{s}}}(B^{n}
|E^{n})_{\sigma^{n}}-\mathrm{leak}_{\text{ec}}-\log_{2}(2/\varepsilon
_{\text{cor}}).
\end{align}
We then replace the above expression in Eq.~\eqref{boundkeyR},
which leads to a stricter upper bound for the key length
\begin{align}
s_{n} & \leq H_{\text{min}}^{\varepsilon_{\text{s}}}(B^{n}|E^{n})_{\sigma^{n}} +2 \log_2(2\varepsilon_{\text{h}}) \nonumber \\
&  -\mathrm{leak}_{\text{ec}}-\log_{2}(2/\varepsilon
_{\text{cor}}) \nonumber \\
&=
H_{\text{min}}^{\varepsilon_{\text{s}}}(B^{n}|E^{n})_{\sigma^{n}}%\log_2(2\varepsilon^{2}_{\text{h}}\varepsilon_{\text{cor}}) 
-\mathrm{leak}_{\text{ec}}+\theta, \label{lenkey_leak}
\end{align}
where we have set
\begin{align}
\theta&:=\log_2(2\varepsilon^{2}_{\text{h}}\varepsilon
_{\text{cor}}).
\end{align}
Note that we include the more precise term $\theta$ instead of just $\log_2(2\varepsilon^{2}_{\text{h}})$ as in past
derivations~\cite{Pirandola21,Pirandola21b,Panos21}.

\subsection{Tensor-product reduction and asymptotic equipartition property}
%By extending a derivation from Ref.~\cite{Pirandola21} 
We may replace the smooth-min entropy of
the sub-normalized state $\sigma^n$ after EC with that of the
normalized state $\rho^{\otimes n}$ before EC. As we show in Appendix~\ref{appEXT}, we
may write the following tensor-product reduction
\begin{equation}
H_{\text{min}}^{\varepsilon_\text{s}}(B^{n}|E^{n})_{\sigma^{n}}
\geq
H_{\text{min}}^{\varepsilon_\text{s}}(B^{n}|E^{n})_{\rho^{\otimes
n}}\label{step_interim}.
\end{equation}
This is a major improvement with respect to Ref.~\cite{Pirandola21}.

Because the state before the EC projection has a tensor product form (under collective attacks),
we can now write a simpler (but larger) lower bound that is based on the von Neumann entropy of the single-copy state $\rho$ in Eq.~\eqref{sharedSSS}.
In fact, we may apply the asymptotic equipartition property
(AEP)~\cite[Cor.~6.5]{TomaThesis} and write
\begin{equation}
H_{\text{min}}^{\varepsilon_\text{s}}(B^{n}|E^{n})_{{\rho}^{\otimes
n}} \geq n H(B|E)_\rho-\sqrt{n} \Delta_\text{aep}, \label{AEPeq}
\end{equation}
where
\begin{align}
\Delta_{\text{aep}}  &  :=4\log_{2}\left(  \sqrt
{|\mathcal{L}|}+2\right)  \sqrt{-\log_{2}\left(  1-\sqrt{1-\varepsilon_{\text{s}}^{2}%
}\right)  }\nonumber\\
&  \simeq4\log_{2}\left(\sqrt{|\mathcal{L}|}+2\right)  \sqrt{\log_{2}(2/\varepsilon
_{\text{s}}^{2})}, \label{AEPd}%
\end{align}
and $|\mathcal{L}| = 2^d$ is the cardinality of the discretized variable $l$ (see Ref.~\cite[Th.~6.4]{TomaThesis} and Ref.~\cite[Sec.~2.F.1]{Pirandola21b}).

\subsection{Upper bound for the secret-key rate\label{ubsec}}
Using Eqs.~\eqref{step_interim} and~\eqref{AEPeq} in
Eq.~\eqref{lenkey_leak}, we may write the following stricter upper
bound
\begin{equation}\label{central_eq}
s_n\leq n H(B|E)_\rho-\text{leak}_\text{ec}-\sqrt{n}\Delta_\text{aep}+\theta,
\end{equation}
where $\rho$ is the single-copy state in Eq.~\eqref{sharedSSS}.
%and
%\begin{align}\theta&=\log_2(2\varepsilon^{2}_{\text{h}}\varepsilon_{\text{cor}}).\end{align}
We finally expand the conditional entropy as
\begin{equation}
H(B|E)_{\rho} = H(l|E)_{\rho}=H(l)-\chi(l:E)_{\rho}, \label{ff1}%
\end{equation}
where $H(l)$ is the Shannon entropy of $l$, and $\chi(l:E)_{\rho}$ is Eve's
Holevo bound with respect to $l$. Therefore, we get
\begin{align}
s_{n} & \leq   n [H(l)- \chi(l:E)_{\rho}] -\mathrm{leak}_{\text{ec}} \nonumber \\
& ~~~~~-\sqrt{n}\Delta_\text{aep} +\theta.
\label{boundkey_general}
\end{align}
Alternatively, this can be written as
\begin{align}
\boxed{ s_{n}\leq n R_{\infty}  -\sqrt{n}\Delta_\text{aep} + \theta,} %\log_2(2 \varepsilon^{2}_{\text{h}}\varepsilon_{\text{cor}}),}
\label{boundkey_final}
\end{align}
where we have introduced the asymptotic key rate
\begin{equation}\label{asym_EC_rate}
     R_{\infty} = H(l)- \chi(l:E)_{\rho} - n^{-1} \mathrm{leak}_{\text{ec}}.
\end{equation}

The result in Eq.~\eqref{boundkey_final} is an upper bound to the number of secret random bits that Alice and Bob can extract with epsilon security $\varepsilon=\varepsilon_{\text{cor}}+\varepsilon_{\text{s}}+\varepsilon
_{\text{h}}$. Note that the secret key rate will need to account for the fact
that this amount of bits is generated with probability $p_{\text{ec}}$ and that only a fraction $n/N$ of the total systems are used for key generation. Thus, the composable secret key rate (bits per use) of a generic
CV-QKD\ protocol under collective attacks is given by
\begin{equation}
R =\frac{p_{\text{ec}}s_{n}}{N}. \label{key_rate_final}
\end{equation}
More explicitly, we have the upper bound
\begin{equation}
\boxed{R \le R_{\text{UB}} = \frac{p_{\text{ec}}[n R_{\infty}  -\sqrt{n}\Delta_\text{aep} + \theta]}{N}.} \label{key_rate_final2_sub}
\end{equation}

\subsection{Achievable key rate for optimal PA~\label{achivable_rate}}
The result in Eq.~\eqref{key_rate_final2_sub} means that Alice and
Bob cannot exceed $R_{\text{UB}}$ bits per use if they want to
have $\varepsilon$-security assured. Assuming they can implement
optimal PA, they can reach a rate $R^{\text{opt}}$ which is still
bounded by $R_{\text{UB}}$ from above, but we can also guarantee
that at least $R_{\text{LB}}$ bits per use are generated.
Basically, for a protocol with optimal extraction of
randomness~\cite[Sec.~8.2]{TomaThesis}, we may have a guaranteed
$\varepsilon$-security and a rate satisfying $R_{\text{LB}} \le
R^{\text{opt}} \le R_{\text{UB}}$, where $R_{\text{UB}}$ is given
in Eq.~\eqref{key_rate_final2_sub} and
\begin{equation}\label{LB}
   \boxed{R_{\text{LB}} = \frac{p_{\text{ec}}[n R_{\infty}  -\sqrt{n}\Delta_\text{aep} +
    \theta-1]}{N}.}
\end{equation}

The lower bound in Eq.~\eqref{LB} is proven by repeating the proof
and using the direct part of the leftover hash
bound~\cite{TomaRenner} (see also Ref.~\cite[Eq.~(8.7)]{TomaThesis})
for the number of bits $s_{n}^{\text{opt}}$ that are achievable by
a protocol with optimal data processing. For this number, we may in fact 
write
\begin{equation}
s_{n}^{\text{opt}}\geq
H_{\text{min}}^{\varepsilon_{\text{s}}}(B^{n}|E^{n})_{\sigma^{n}}+2
\log_2(\sqrt{2}\varepsilon_{\text{h}}). \label{boundkey_LB}
\end{equation}
%together with the converse part stated in Eq.~\eqref{boundkey}.
We can see that the $-1$ difference between
Eqs.~\eqref{boundkey} and~\eqref{boundkey_LB} become an extra
$-p_\text{ec}/N$ in Eq.~\eqref{LB}. Because $N$ is typically
large, we also see that $R_{\text{LB}} \simeq R_{\text{UB}}$.
%which means that both quantities can be used to estimate the key rate.

Note that the direct leftover hash bound was used in the
derivations of Refs.~\cite{Pirandola21,Pirandola21b,Panos21},
which therefore provided formulas for the rate achievable by protocols
with optimal PA. However, these previous
works are more pessimistic than our current result due to a different tensor-product reduction with respect to Eq.~\eqref{step_interim}. In particular, the
key-rate lower bound from Ref.~\cite{Pirandola21b} takes the form
\begin{equation}\label{LB_OLD}
    R_{\text{LB}}^{\text{old}} = \frac{p_{\text{ec}}[n R_{\infty}  -\sqrt{n}\Delta'_\text{aep} +
    \theta'-1]}{N},
\end{equation}
where $\theta' = \theta + \log_2 [p_\text{ec}
(1-\varepsilon^{2}_{\text{s}}/3)]$, and
\begin{equation}
\Delta'_\text{aep} = [\Delta_\text{aep}]_{\varepsilon_\text{s}
\rightarrow p_\text{ec} \varepsilon^{2}_\text{s}/3} .
\end{equation}
(To be precise the formula above is already a refinement since we
have also included more precise leakage contribution, as explained
in Sec.~\ref{sec_leak}).

\subsection{Specification to various protocols\label{spec_prot}}
\subsubsection{Formula for discrete-alphabet coherent state protocols}
More specific formulas for a discrete-alphabet protocol are
immediately derived. Let us define the reconciliation parameter
$\beta\in\lbrack0,1]$ by setting
\begin{equation}
H(l)-n^{-1} \mathrm{leak}_{\text{ec}} =\beta I(k:l), \label{ff2}
\end{equation}
where  $I(k:l)$ is Alice and Bob's mutual information. Then, the asymptotic key rate takes the form
\begin{equation}
R_{\infty}=\beta I(k:l)-\chi(l:E)_{\rho}. \label{asy1}
\end{equation}
This is to be replaced in  Eq.~\eqref{key_rate_final2_sub} for the upper bound, and Eq.~\eqref{LB} for the lower bound with optimal PA.

\subsubsection{Formula for Gaussian-modulated coherent state protocols}
In the case of a Gaussian-modulated protocol, we need to express the formulas in terms of quadratures. First, we re-define the reconciliation parameter $\beta\in\lbrack0,1]$ as
\begin{equation}
H(l)-n^{-1} \mathrm{leak}_{\text{ec}} =\beta I(x:y), \label{ff2bis}%
\end{equation}
where $I(x:y)\geq I(k:l)$ is Alice and Bob's mutual information
computed over their continuous variables. Second, we exploit the data
processing inequality for Eve's Holevo bound, so
$\chi(l:E)_{\rho}\leq\chi(y:E)_{\rho}$\ under digitalization
$y\overset{\text{ADC}}{\rightarrow}l$. Thus, we can use the
asymptotic rate \begin{equation} R_{\infty}=\beta I(x:y)-
\chi(y:E)_{\rho},\label{asy2}
\end{equation}
to be replaced in the previous general formulas.

\subsubsection{Other protocols}
Other protocols can be considered. For example, the composable key
rate of CV-MDI-QKD can be expressed using our general formulation
once we replace the corresponding asymptotic expression
$R_{\infty}$. The same can be stated for post-selection protocols,
which also involves the introduction of an extra (post-selection)
probability $p_\text{ps}$, appearing as a further pre-factor in
Eqs.~\eqref{key_rate_final2_sub} and~\eqref{LB}, i.e.,
$p_{\text{ec}} [\dots]/N \rightarrow p_{\text{ps}} p_{\text{ec}}
[\dots]/N$. In general, the post-selection process can be seen as
a global filter that distills the number of runs and is applied
before the standard processing of data via EC and PA.

\subsection{Parameter estimation\label{PE_sec}}
The asymptotic key rate $R_{\infty}$ depends on a number
$n_{\text{pm}}$ of parameters $\mathbf{p}$. By sacrificing $m$
systems, Alice and Bob can compute maximum likelihood estimators
$\mathbf{\hat{p}}$ and worst-case values $\mathbf{p}_{\text{wc}}$,
which are $w$ standard deviations away from the mean values of the
estimators. The worst-case value bounds the true value of a
parameter up to an error probability $\varepsilon
_{\text{pe}}=\varepsilon_{\text{pe}}(w)$. This means that, overall, $n_{\text{pm}}$ worst-case values $\mathbf{p}_{\text{wc}}$\ will bound the parameters
$\mathbf{p}$\ up to a total error probability $\simeq n_{\text{pm}}%
\varepsilon_{\text{pe}}$. Because PE occurs before EC, this probability needs to be multiplied by $p_\mathrm{ec}$, so we have a total modified epsilon security
\begin{equation}
\varepsilon = \varepsilon_{\text{cor}}+\varepsilon_{\text{s}}+\varepsilon
_{\text{h}}+p_{\text{ec}}n_{\text{pm}}\varepsilon_{\text{pe}}.
\end{equation}

In the composable formulas of Eqs.~\eqref{key_rate_final2_sub}
and~\eqref{LB}, the asymptotic term $R_{\infty} = R_{\infty}
(\mathbf{p})$ will be computed on the estimators and worst-case
values, i.e., replaced by
\begin{equation}
    R_{\infty}^{\text{pe}}:=R_{\infty}(\mathbf{\hat{p}},\mathbf{p}_{\text{wc}}).\label{PEadaptation}
\end{equation}
In particular, the expressions in Eqs.~\eqref{asy1}
and~\eqref{asy2} will be replaced by
\begin{equation}
 R_{\infty}^{\text{pe}}=\beta [I]_{\mathbf{\hat{p}}}-[\chi_{\rho}]_{\mathbf{p}_{\text{wc}}}.\label{asy3}
\end{equation}

\subsection{From one block to a session of blocks\label{qkd-session}}
In a typical fiber-based scenario, a QKD session is stable, i.e.,
the main channel parameters are constant for a substantial period
of time. This means that we can consider a session of
$n_{\text{bks}}$ blocks, each block with size $N$. In this
scenario, the success probability $p_{\text{ec}}$ becomes the
fraction of blocks that survive EC (the value $1-p_{\text{ec}}$
is also known as frame error rate). Assuming such a stable QKD
session, PE can be performed on a large number of points, namely
$n_{\text{bks}} m$. This approach leads to better estimators and
worst-case values to be used in Eq.~\eqref{PEadaptation}. Using these
improved statistics, Alice and Bob will then implement EC
block-by-block. Each block surviving EC will undergo PA, where it
is subject to a hash function randomly chosen from a $2$-universal
family. Each block compressed by PA is then concatenated into the
final key.

\subsection{Extension to coherent attacks for heterodyne\label{Coh_att}}
One can extend the security of the Gaussian-modulated protocol with heterodyne detection to coherent attacks, following the
Gaussian de Finetti reduction of Ref.~\cite{Lev2017}. %This protocol signal states display a symmetry under the unitary group that can be exploited to apply a Gaussian de Finetti reduction achieving a security level against more powerful attacks~\cite{Lev2017}. For such a treatment to be valid, 
The parties need to verify that the Hilbert space of the signal states is suitably constrained. In other words, the energy of Alice's and Bob's states should be less than some threshold values, $d_A$ and $d_B$, respectively.  The parties execute a random energy test over $k$ states to estimate the energy of the other $n$ signal states that participate in the standard steps of the protocol.  Given that the test is successful with probability $p_\text{en}$ and that the protocol is $\varepsilon$-secure against collective Gaussian attacks, the new key length is decreased by the following amount of secret bits~\cite{Lev2017}
$s^\prime_{n}\leq s_{n}-\Phi$,
where
\begin{equation}
\Phi := 2\left \lceil \log_2 \binom{K+4}{4}\right \rceil,
\end{equation}
and
\begin{equation}
K=\max\left\{1,n(d_A+d_B)\frac{1+2\sqrt{\frac{\ln(8/\varepsilon)}{2n}}+\frac{\ln(8/\varepsilon)}{n}}{1-2\sqrt{\frac{\ln(8/\varepsilon)}{2k}}}\right\}.
\end{equation}
The number of channel uses per block is extended to $N^\prime=N+k$, the epsilon-security is rescaled to
\begin{equation}\label{epsilon_coherent}
\varepsilon^\prime=\frac{K^4}{50} \varepsilon,
\end{equation}
and the probability of not aborting $p_\text{ec}$ is replaced by $p_\text{ec}\rightarrow p_\text{en}p_\text{ec}$. %in Eq.~\eqref{key_rate_final2_sub}. 
One may set Alice's energy threshold to be larger than the mean photon number $\bar{n}_A=V/2$ of the average thermal state created by her classical modulation $V$. %(see also Eq.~\eqref{mutual_info_gen}). 
More specifically, taking into account statistical calculations due to the use of $k$ signal states, one may set $d_A\geq\bar{n}_A+\mathcal{O}(k^{-1/2})$. Then, under the assumption of a lossy channel with reasonable excess noise,
the mean number of photons received by Bob is smaller than $\bar{n}_A$, so if we set $d_B=d_A$, we certainly have $d_B\geq\bar{n}_A+\mathcal{O}(k^{-1/2})$. These conditions lead to an almost successful energy test $p_\text{en}\simeq 1$. Consequently, the secret key rate of the heterodyne protocol under coherent attacks will be given by
\begin{equation}
R^\prime=\frac{p_\text{ec}s^\prime_n}{N^\prime},
\end{equation}
constrained by the upper bound [similar to Eq.~\eqref{key_rate_final2_sub}]
\begin{equation}
R^\prime \le R^\prime_{\text{UB}} = \frac{p_{\text{ec}}[n R_{\infty}  -\sqrt{n}\Delta_\text{aep} + \theta-\Phi]}{N^\prime},\label{key_rate_final3_sub}
\end{equation}
and the lower bound [similar to Eq.~\eqref{LB}]
\begin{equation}
R^\prime \ge R^\prime_{\text{LB}} = R^\prime_{\text{UB}} - \frac{p_{\text{ec}}}{N^\prime}. \label{key_rate_final3_sub_lb}
\end{equation}

\subsection{Practical Considerations\label{pract_sect}}
In an experimental implementation of a CV-QKD protocol, the parties have to numerically estimate two crucial parameters: the EC probability $p_\text{ec}$ and the reconciliation efficiency $\beta$. The EC probability can be computed as the ratio $p_\text{ec}=\frac{n_\text{ec}}{n_\text{bks}}$ between the $n_\text{ec}$ successfully corrected blocks and the total number of blocks of a session $n_\text{bks}$, assuming that the channel is stable (see also~\cite{Alex_I,Alex_II,Alex_III}). The reconciliation efficiency can be computed from the leakage of the EC scheme employed. Typically, the EC scheme exploits non-binary LDPC codes, described by a $c \times n$ parity check matrix with code rate $R_\text{code}= c/n$, where $c$ is the number of parity checks. In this case, the leakage can be bounded by
\begin{equation}
n^{-1}\text{leak}_\text{ec}\leq d_{\mathrm{least}}-R_\text{code}d_{\mathrm{syn}},
\end{equation}
where $d_{\mathrm{least}}$ is the number of the least significant bits sent on the clear, while $d_{\mathrm{syn}}$ is the number of syndrome bits (see Refs.~\cite{Alex_I, Alex_II} for details and precise definitions). 

Once the leakage is bounded, one may use Eqs.~\eqref{ff2} or \eqref{ff2bis} to compute the reconciliation parameter $\beta$. However, in a practical setting, the value of the entropy $H(l)$ is also not exactly known and must be estimated. During PE, the parties calculate the frequency $f_l=n_l/n$ of the value $l$, starting from its $n_l$ occurrences in the sequence of length $n$. In this way, they construct the estimator   
\begin{equation}\label{ent_est}
\widehat{H}(l)=-\sum_{l=0}^{2^d-1} f_{l}\log_2f_{l}.
\end{equation}
The value of this estimator is then used in Eqs.~\eqref{ff2} or \eqref{ff2bis} to derive an estimate for $\beta$~\cite{endnote2}.

The uncertainty on the value of Bob's entropy has also an effect at the level of the composable key rate, introducing a further epsilon parameter. For the entropy estimator, we have
\begin{equation}
H(l)\geq \mathbb{E}(\widehat{H}(l))\label{ii},
\end{equation}
and we can write
\begin{equation}
\text{Prob}\left[|\widehat{H}(l)-\mathbb{E}(\widehat{H}(l))|\geq\delta_\text{ent}\right]\leq \varepsilon_\text{ent},
\end{equation}
for
\begin{equation}
\delta_\text{ent}=\log_2(n)\sqrt{\frac{2\ln(2/\varepsilon_\text{ent})}{n}}.
\end{equation}
% Equivalently, we can write
% \begin{equation}
% \text{Prob}\left[|\widehat{H}(l)-\mathbb{E}(\widehat{H}(l))|\leq \delta_\text{ent}\right]\geq 1-\varepsilon_\text{ent},
% \end{equation}
% meaning that we have the condition
This means that we have the condition
\begin{align}
&-\delta_\text{ent} \leq \widehat{H}(l)-\mathbb{E}(\widehat{H}(l))\leq \delta_\text{ent}
\end{align}
with probability larger than $1-\varepsilon_\text{ent}$.

Combining the inequality above with Eq.~\eqref{ii}, we get
\begin{align}
%&\widehat{H}(l)-H(l)\leq \delta_\text{ent}\\
%&H(l)-\widehat{H}(l)\geq -\delta_\text{ent}\\
&H(l)\geq\widehat{H}(l) -\delta_\text{ent}\label{ent_lb}
\end{align}
up to an error probability $\varepsilon_\text{ent}$. 
In other words, we can replace Bob's entropy in the asymptotic rate of Eq.~\eqref{asym_EC_rate} with the lower bound in Eq.~\eqref{ent_lb} computed from the estimator in Eq.~\eqref{ent_est}. This leads to a stricter upper bound for the composable secret key rate. More precisely, Eq.~\eqref{boundkey_final} becomes 
\begin{align}
s_{n}\leq n \widehat{R}_{\infty} -n \delta_\text{ent} -\sqrt{n}\Delta_\text{aep} + \theta,
%\boxed{ \widehat{s}_{n}\leq n \widehat{R}_{\infty} -n \delta_\text{ent} -\sqrt{n}\Delta_\text{aep} + \theta,} %\log_2(2 \varepsilon^{2}_{\text{h}}\varepsilon_{\text{cor}}),}
\label{boundkey_final_hat}
\end{align}
where the asymptotic key rate becomes
\begin{equation}\label{asym_EC_rate_hat}
     \widehat{R}_{\infty} = \widehat{H}(l)- \chi(l:E)_{\rho} - n^{-1} \mathrm{leak}_{\text{ec}}.
\end{equation}
Thus, the corresponding composable secret key rate 
\begin{equation}
R=\frac{p_{\text{ec}}s_{n}}{N} \label{app_key_rate_final_hat}
\end{equation}
is  upper-bounded by
\begin{equation}
R \le \widehat{R}_{\text{UB}} = \frac{p_{\text{ec}}[n \widehat{R}_{\infty}  - n \delta_\text{ent}-\sqrt{n}\Delta_\text{aep} + \theta]}{N}, \label{key_rate_final2_sub_hat}
\end{equation}
with overall $\varepsilon$-security $\varepsilon=\varepsilon_{\text{cor}}+\varepsilon_{\text{s}}+\varepsilon
_{\text{h}}+p_\text{ec}\varepsilon_\text{ent}$. Note that $\varepsilon_\text{ent}$ is re-scaled by $p_\text{ec}$ because Bob's entropy is evaluated during PE and, therefore, before EC. Similarly, according to the discussion in Sec.~\ref{achivable_rate}, we may write the lower bound
\begin{equation}\label{appLB}
R^\text{opt}\geq  \widehat{R}_{\text{UB}}-\frac{p_\text{ec}}{N} 
\end{equation}
for a protocol with optimal PA.
% \begin{equation}\label{appLB}
%    \boxed{\widehat{R}_{\text{LB}} = \frac{p_{\text{ec}}[n \widehat{R}_{\infty} -n\delta_\text{ent} -\sqrt{n}\Delta_\text{aep} +
%     \theta-1]}{N}.}
% \end{equation}

Including the estimation of the channel parameter $\mathbf{p}$ via $\widehat{\mathbf{p}}$ and $\mathbf{p}_\text{wc}$, the asymptotic rate in Eq.~\eqref{asym_EC_rate_hat} becomes $\widehat{R}^\text{pe}_\infty=\widehat{R}_\infty(\widehat{\mathbf{p}},\mathbf{p}_\text{wc})$. In particular for the protocols in Sec.~\ref{spec_prot}, the rates in Eqs.~\eqref{asy1} and~\eqref{asy2} become
 
%\begin{equation}
%Eq.~\eqref{ff2} takes the following form
%\begin{equation}\label{ff2_hat}
%\widehat{H}(l)-n^{-1} \mathrm{leak}_{\text{ec}} =\widehat{\beta} [I(k:l)]_{\hat{\mathbf{p}}}, 
%\end{equation}
\begin{equation}
\widehat{R}^{\text{pe}}_{\infty}=\widehat{\beta} [I]_{\hat{\mathbf{p}}}-[\chi_\rho]_{\mathbf{p}_\text{wc}}.\label{asy1_hat}
\end{equation}
% where $\widehat{\beta}$ is computed according to Eq.~\eqref{ff2} or~\eqref{ff2bis} up to replacing $H(l)$ with $\widehat{H}(l)$ and $I$ with $[I]_{\hat{\mathbf{p}}}$.
By replacing $\widehat{R}^\text{PE}_\infty\rightarrow\widehat{R}_\infty$ in Eq.~\eqref{key_rate_final2_sub_hat}, we therefore bound the composable secret key rate, which accounts for the entire PE process,
%\begin{equation}
%\widehat{R}^{\text{pe}} =\frac{p_{\text{ec}}\widehat{s}^{\text{pe}}_{n}}{N}. \label{key_rate_final_hat}
%\end{equation}
% This is upper-bounded by
% \begin{equation}
% \boxed{\widehat{R}^{\text{pe}} \le \widehat{R}^{\text{pe}}_{\text{UB}} = \frac{p_{\text{ec}}[n \widehat{R}_{\infty}^{\text{pe}}  - n \delta_\text{ent}-\sqrt{n}\Delta_\text{aep} + \theta]}{N}} \label{key_rate_final2_sub_hat}
% \end{equation}
with overall $\varepsilon$-security 
\begin{equation}
    \varepsilon=\varepsilon_{\text{cor}}+\varepsilon_{\text{s}}+\varepsilon
_{\text{h}}+p_\text{ec} \varepsilon_\text{ent}+p_{\text{ec}}n_{\text{pm}}\varepsilon_{\text{pe}}.\label{eps_sec_total}
\end{equation}
Finally, for the heterodyne protocol, we may extend the security to coherent attacks repeating the modifications that lead to Eqs.~\eqref{key_rate_final3_sub} and~\eqref{key_rate_final3_sub_lb} of Sec.~\ref{Coh_att}.

\section{Examples with the main Gaussian-modulated protocols\label{SEC3}}
In order to use the composable formula, we need to specify the asymptotic key rate and the PE procedure, so that we can compute the rate $R_{\infty}^{\text{pe}}$ in Eq.~\eqref{PEadaptation} to be replaced in Eq.~\eqref{key_rate_final2_sub}. Here, we report the known formulas for the asymptotic key rates of the Gaussian-modulated coherent-state protocols (with homodyne and heterodyne detection). These asymptotic formulas can be found in a number of papers (e.g., see Ref.~\cite{AOP} and references therein). Then we consider the modifications due to PE. 

\subsection{Gaussian modulation of coherent states with homodyne detection}
We model the link connecting the parties as a thermal-loss channel with transmissivity $T=10^{-D/10}$ (where $D$ is here the loss in dB) and excess noise $\xi$. The dilation of the channel is represented by a beam splitter with transmissivity $T$ that Eve uses to inject one mode of a two-mode squeezed vacuum (TMSV) state with variance
\begin{equation}\label{eq:omega}
\omega=\frac{T\xi}{1-T}+1.
\end{equation}
Eve's injected mode is therefore coupled with Alice's incoming mode via the beam splitter and the output is received by Bob, who detects it using a homodyne detector with efficiency $\eta$ and electronic noise $u_\text{el}$ (both local parameters that can be considered to be trusted in a well-calibrated scenario). The other, environmental, output of the beam splitter is stored by Eve in a quantum memory, together with the kept mode of the TMSV state. In this way, many modes are collected in Eve's quantum memory, which is finally subject to an optimal joint measurement (collective entangling-cloner attack).  

Alice and Bob's mutual information is given by
\begin{align}
I(x:y)%=&H(y)-H(y|x)=\frac{1}{2}\log_2\left(\frac{\sigma_y^2}{\sigma_{z}^2}\right)\notag\\
=&\frac{1}{2}\log_2 \left[ 1+\frac{V}{\xi+(1+u_\text{el})/(T\eta)} \right], \label{mutual_info_homo}
\end{align} 
where $V$ is Alice's modulation.
The CM of Eve's output state (her partially-transmitted TMSV state) is given by 
\begin{equation}\label{eq:totCM}
\mathbf{V}_{E}=\begin{pmatrix}
\omega\mathbf{I}&\psi \mathbf{Z}\\
\psi\mathbf{Z}&\phi\mathbf{I}
\end{pmatrix},
\end{equation}
where $\mathbf{I}=\text{diag}(1,1)$, $\mathbf{Z}=\text{diag}(1,-1)$ and
\begin{align}
\psi=&\sqrt{T(\omega^2-1)},\\
\phi=&T\omega+(1-T)(V+1).
\end{align}
Then, Eve's conditional CM (conditioned on Bob's outcome) is given by
\begin{equation}
\mathbf{V}_{E|y}=\mathbf{V}_{E}-b^{-1}\begin{pmatrix}
\gamma^2 \boldsymbol{\Pi}&\gamma \theta \boldsymbol{\Pi}\\
\gamma \theta \boldsymbol{\Pi} &\theta^2 \boldsymbol{\Pi}
\end{pmatrix},
\end{equation}
where $\boldsymbol{\Pi}=\text{diag}\{1,0\}$ and
\begin{align}
b=& T \eta(V+\xi)+1+u_\text{el},\label{beta}\\
\gamma=&\sqrt{\eta(1-T)(\omega^2-1)},\label{gamma}\\
\theta=&\sqrt{\eta T(1-T)}(\omega-V-1)\label{theta}.
\end{align}
By calculating the symplectic eigenvalues of the total CM, $\nu_+$ and $\nu_-$, and those of the conditional CM, $\tilde{\nu_+}$ and $\tilde{\nu}_-$, we obtain Eve's Holevo information on Bob's outcome
\begin{equation}
\chi(E:y)=h(\nu_+)+h(\nu_-)-h(\tilde{\nu}_+)-h(\tilde{\nu}_-),
\end{equation}
where we use the usual CV-based entropy function
\begin{equation}
h(\nu):=\frac{\nu+1}{2}\log_2\frac{\nu+1}{2}-\frac{\nu-1}{2}\log_2\frac{\nu-1}{2}.
\end{equation}
Then the asymptotic secret key rate is given by the difference between the mutual information (multiplied by the reconciliation efficiency $\beta$) and Eve's Holevo information as in Eq.~\eqref{asy2}. 
\subsection{Gaussian modulation of coherent states with heterodyne detection}
For the protocol with heterodyne detection, the mutual information is a simple modification of the previous one in Eq.~\eqref{mutual_info_homo} and given by
% \begin{align}
% I(x:y)%=&H(y)-H(y|x)=\frac{1}{2}\log_2\left(\frac{\sigma_y^2}{\sigma_{z}^2}\right)\notag\\
% =&\log_2 \left[ 1+\frac{V}{\xi+(2+u_\text{el})/(T\eta)} \right].\label{mutual_info_het}
% \end{align} 
\begin{equation}\label{mutual_info_gen}
I(x:y)=\frac{V_0}{2}\log_2 \left[ 1+\frac{V}{\xi+(V_0+u_\text{el})/(\eta T)} \right],
\end{equation}
where $V_0 = 2$ (note that for $V_0 = 1$ we get the expression valid for homodyne detection).
Eve's CM is the same as in Eq.~\eqref{eq:totCM}, but the conditional CM is instead given by
\begin{equation}
\mathbf{V}_{E|y}=\mathbf{V}_{E}-(b+1)^{-1}\begin{pmatrix}
\gamma^2 \mathbf{I}&\gamma \theta\mathbf{Z}\\
\gamma \theta\mathbf{Z} &\theta^2 \mathbf{I}
\end{pmatrix}.
\end{equation}
\begin{figure}[b]
\vspace{-0.45cm}
\centering
\includegraphics[width=0.45\textwidth]{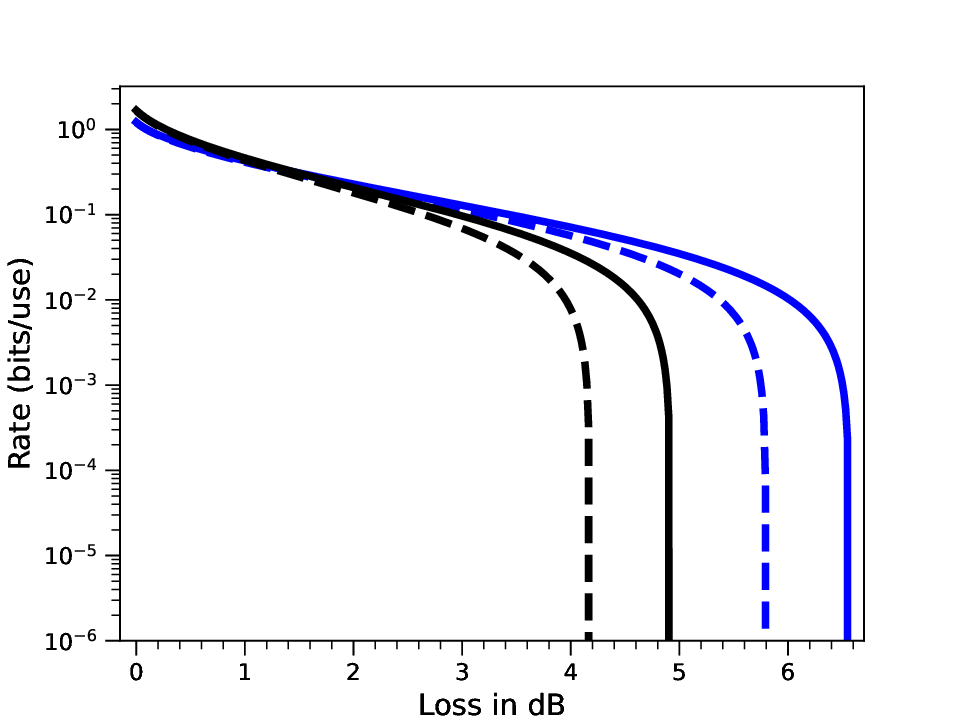}
\caption{Improved composable secret key rate [upper bound of Eq.~\eqref{key_rate_final2_sub}] for the Gaussian modulated coherent-state protocol with homodyne detection (blue solid line) and heterodyne detection (black solid line) with respect to channel loss in dB. These lines overlap with those associated with the lower bound of Eq.~\eqref{LB}. The corresponding dashed lines are computed using Eq.~\eqref{LB_OLD}, based on previous literature. We have set $\beta = 0.98$ and $p_\text{ec} = 0.95$. Excess noise is $\xi = 0.01$, detection efficiency is $\eta = 0.6$, and electronic noise is $ u_\text{el} = 0.1$. Security epsilons have all been set to $2^{-32}$. The cardinality of the alphabet is $|\mathcal{L}| = 2^7$ for homodyne and $|\mathcal{L}|= 2^{14}$ for heterodyne. Block size is $N = 10^7$ and PE is based on $m = N/10$ sacrificed signals. We have optimized the results over the variance $V$ of Alice's Gaussian modulation.}
\label{fig:fig_improved}
\end{figure}
\begin{figure}[t]
\vspace{-0.45cm}
\centering
\includegraphics[width=0.45\textwidth]{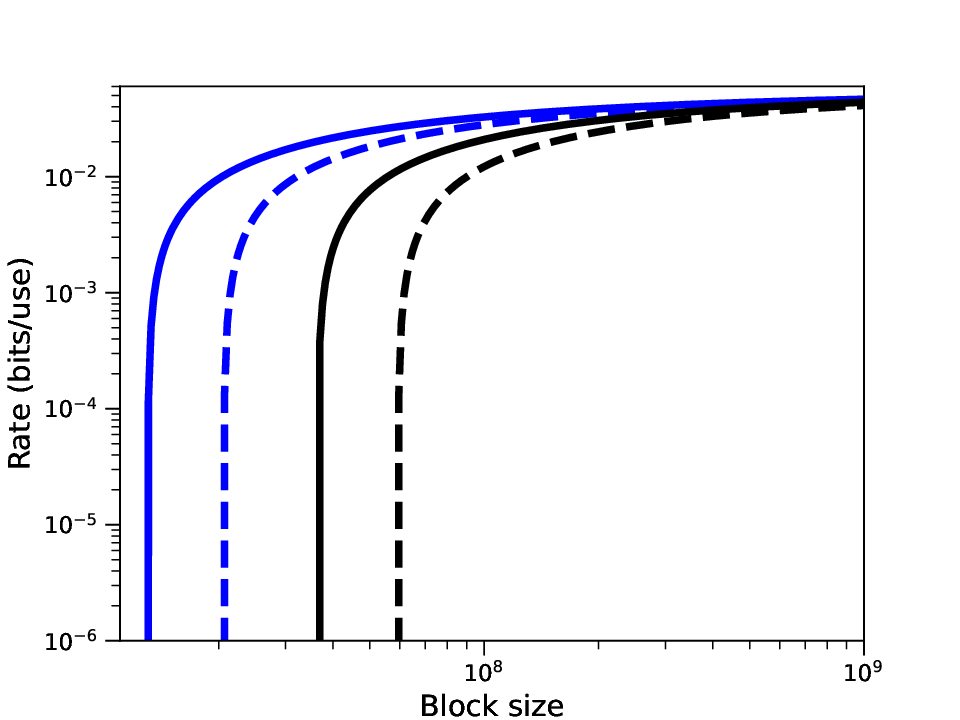}
\caption{Improved composable secret key rate [upper bound of Eq.~\eqref{key_rate_final2_sub}] for the Gaussian modulated coherent-state protocol with homodyne detection (blue solid line) and heterodyne detection (red solid line) with respect to the block size $N$. These lines coincide with those computed from the lower bound of Eq.~\eqref{LB}. The corresponding dashed lines are computed using Eq.~\eqref{LB_OLD}, based on previous literature. Loss is set to $7$ dB, while all the other parameters are chosen as in Fig.~\ref{fig:fig_improved}.}
\label{fig:fig_bszk_improved}
\end{figure}

\subsection{Parameter estimation and final performance}
Let us now include PE, assuming that $m$ signals are sacrificed for building the estimators of the channel parameters (to be used in the mutual information) and the associated worst-case values (to be used in Eve's Holevo bound). One therefore computes estimators $\hat{T} \simeq T$, $\hat{\xi} \simeq \xi$ and the following worst-case values
\begin{align}
T_\text{wc}& \simeq T-w \sigma_T,\\
\xi_\text{wc}&\simeq
\frac{T}{T_\text{wc}}\xi+w \sigma_{\xi},
%T_\text{wc}& \simeq T-\frac{2wT}{\sqrt{V_0 m}} \sqrt{c_{\mathrm{PE}}+ \frac{\xi+\frac{V_0+u_\text{el}}{\eta T}}{V}},\label{TmethI}\\
%\xi_\text{wc}& \simeq \frac{\eta T \xi+\frac{\sqrt{2} w}{\sqrt{V_0 m}} %\left(\eta T \xi +V_0+u_\text{el}\right)}{\eta T_\text{wc}},\label{ximethI}
\end{align}
where %$w$ is chosen as in Eq.~\eqref{PE-w}.
\begin{align}
\sigma_T&=\frac{2T}{\sqrt{V_0 m}}\sqrt{c_\text{pe}+\frac{\xi+\frac{V_0+u_\text{el}}{\eta T}}{V}},\label{sigma_tau}\\
\sigma_\xi&=\sqrt{\frac{2}{V_0 m}}\frac{\eta T\xi+V_0+u_\text{el}}{\eta T_\text{wc}}.\label{sigma_xi}
\end{align}

In the equations above, $V_0=1$ is for homodyne detection and $V_0=2$ is for heterodyne detection. 
Then, in Eq.~\eqref{sigma_tau}, the term $c_{\mathrm{pe}}$ can be set to zero~\cite{Lev2010} (in fact, another choice would be $c_{\mathrm{pe}}=2$~\cite{Vlad14} based on a weaker assumption~\cite{endnote1}). The parameter $w$ that connects the worst-case values with $\varepsilon_{\mathrm{pe}}$ is simply given by an inverse error function when we assume a Gaussian approximation for the parameters, i.e.,
\begin{equation}
    w=\sqrt{2}\text{erf}^{-1}(1-\varepsilon_\text{pe}).
\end{equation}
However, when stricter conditions are required, e.g., in the case of coherent attacks [see Eq.~\eqref{epsilon_coherent}], we use chi-squared distribution tail bounds where $w$ is given by 
\begin{equation}
    w=\sqrt{2\ln \varepsilon_\text{pe}^{-1}}.
\end{equation}
(See Appendix~\ref{PEtailbounds} for more details.)

Thus, by using $\hat{\bold{p}} = (\hat{T},\hat{\xi})$ and $\bold{p}_\text{wc}=(T_\text{wc},\xi_\text{wc})$, we compute the PE rate as in Eq.~\eqref{asy3} to be replaced in Eq.~\eqref{key_rate_final2_sub} for both the homodyne and heterodyne protocols. 
Let us assume ad hoc values for $p_\text{ec}$ and $\beta$ (the exact numerical values of these parameters are known after a realistic implementation or simulation of EC, as discussed in Sec.~\ref{pract_sect}). Then, we show the performances of the two protocols in Figs.~\ref{fig:fig_improved} and~\ref{fig:fig_bszk_improved}. More specifically, in Fig.~\ref{fig:fig_improved}, we depict the secret key rate versus channel loss, while, in Fig.~\ref{fig:fig_bszk_improved}, we show its behavior with respect to block size. For the sake of comparison, we have also included the results based on previous literature~\cite{Pirandola21, Pirandola21b} (refined in Sec.~\ref{achivable_rate}).  From the figures, we can see a significant improvement in the key rate performance both in terms of robustness to loss and smaller block size.

\begin{figure}[t]
\vspace{-0.0cm}
\centering
\includegraphics[width=0.6\textwidth]{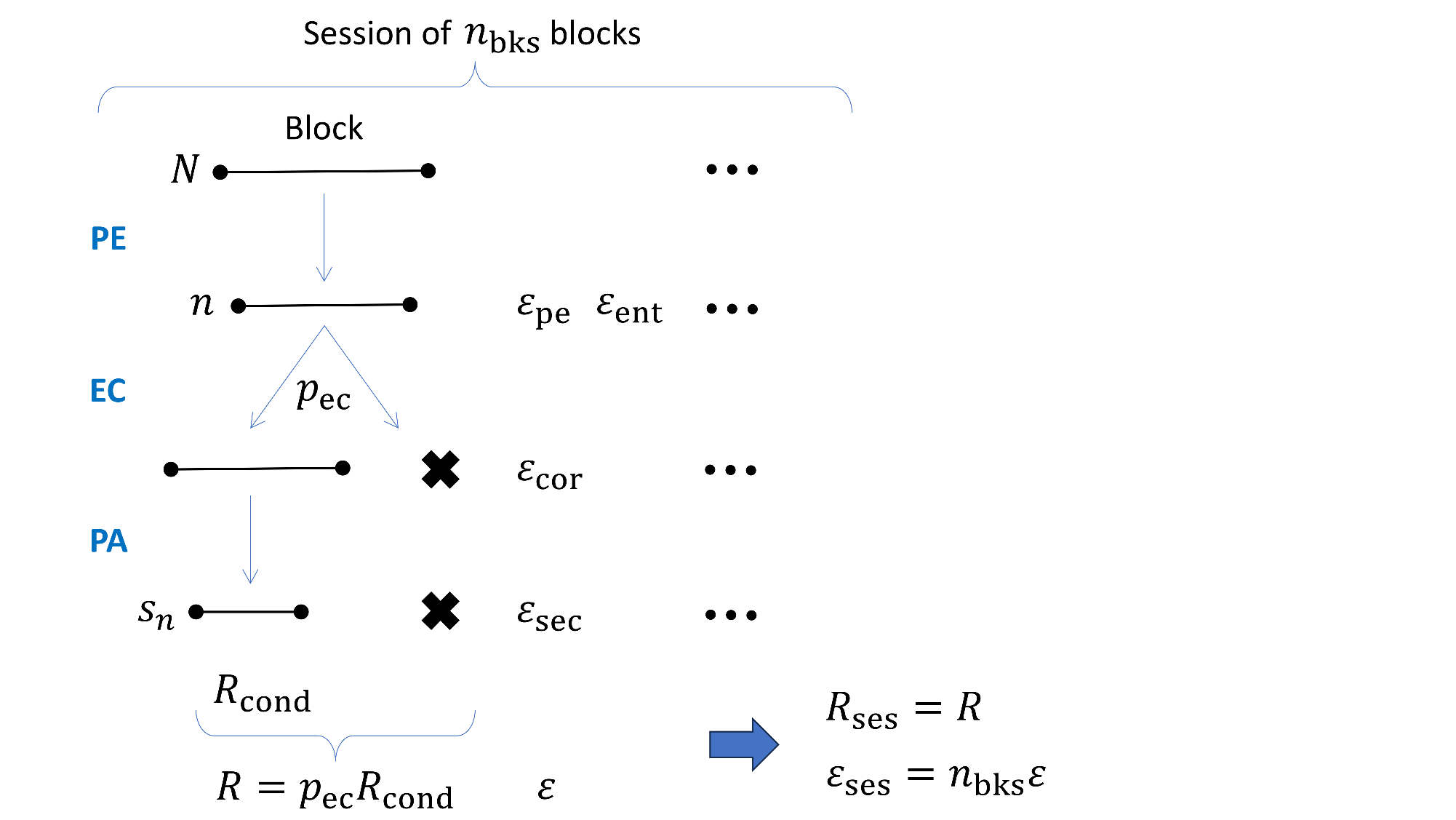}
\caption{Rate and epsilon security, from block to session.}
\label{fig:QKDsession}
\end{figure}

\SP{\section{Discussion\label{sec:clarifications}}}
Let us clarify some important points about the epsilon security and the secret key rate. As previously discussed, under stable channel conditions, one can consider a QKD session of $n_{\mathrm{bks}}$ blocks. Each block of size $N$ undergoes the three subsequent processes depicted in Fig.~\ref{fig:QKDsession}. These are the following:
\begin{itemize}
    \item PE, affected by errors $\varepsilon_{\mathrm{pe}}$ and $\varepsilon_{\mathrm{ent}}$;
    \item EC, performed with success probability $p_{\mathrm{ec}}$, and affected by error $\varepsilon_{\mathrm{cor}}$;
    \item PA, affected by error $\varepsilon_{\mathrm{sec}}$.
\end{itemize}

The PE errors $\varepsilon_{\mathrm{pe}}$ and $\varepsilon_{\mathrm{ent}}$ are introduced before EC, so they will appear in the final epsilon security $\varepsilon$ with probability $p_{\mathrm{ec}}$ (with probability $1-p_{\mathrm{ec}}$ the block is not processed and therefore there is no forward-propagated PE error). Thus, on average, the two types of PE errors contribute as $p_{\mathrm{ec}}\varepsilon_{\mathrm{pe}}$ and $p_{\mathrm{ent}}\varepsilon_{\mathrm{ent}}$ in the final $\varepsilon$.

Situation is different for the correctness $\varepsilon_{\mathrm{cor}}$. This is related to the joint probability that the verification step of EC succeeds and the strings are different [cf. Eq.~\eqref{ecorrectness}]. This means that it is already defined as an average (unconditional) quantity.
In fact, we may write that
\begin{align}
    \mathrm{Prob} (T_\text{ec}\mathrm{~and~}\hat{\mathbf{l}} \neq \mathbf{l}  )=
    p_{\text{ec}}\mathrm{Prob}(\hat{\mathbf{l}} \neq \mathbf{l} \vert T_\text{ec})  \leq \varepsilon_{\text{cor}}
\end{align}
is equivalent to the unconditional probability
\begin{align}
   & \mathrm{Prob}(\hat{\mathbf{l}} \neq \mathbf{l}) = p_{\text{ec}}\mathrm{Prob}(\hat{\mathbf{l}} \neq \mathbf{l} \vert T_\text{ec}) \nonumber \\
   &  +(1-p_{\text{ec}}) \mathrm{Prob}(\hat{\mathbf{l}} \neq \mathbf{l} \vert \mathrm{not ~}T_\text{ec}) \leq \varepsilon_{\text{cor}}.
\end{align}

Similarly, the secrecy $\varepsilon_{\mathrm{sec}}$ is the joint probability that the protocol succeeds (i.e., passes EC) and the output key is not secure [cf. Eq.~\eqref{esecrecy}]. Therefore, this is also an average (unconditional) quantity that contributes directly as is to the epsilon security $\varepsilon$. 

For this reason, the epsilon security takes the form of Eq.~\eqref{eps_sec_total}, i.e.,
\begin{equation}
\varepsilon=\varepsilon_{\text{cor}}+\varepsilon_{\text{sec}}+p_\text{ec} \varepsilon_\text{ent}+p_{\text{ec}}n_{\text{pm}}\varepsilon_{\text{pe}}.\label{eps_sec_total2}
\end{equation}
However, this expression assumes the knowledge of $p_{\mathrm{ec}}$, so it is preferable to use the simpler bound 
\begin{equation}
\varepsilon \le \varepsilon_{\text{cor}}+\varepsilon_{\text{sec}}+ \varepsilon_\text{ent}+n_{\text{pm}}\varepsilon_{\text{pe}}.\label{eps_sec_total3}
\end{equation}
For the session, the total epsilon security is $\varepsilon_{\mathrm{ses}}=n_{\mathrm{nks}} \varepsilon$.

%%%%%%%%%%%%%%%%%%%%%%%%
% Note that we can write Eq.~\eqref{ecorrectness} as
% \begin{equation}
% p_{\text{ec}}\mathrm{Prob}(\hat{\mathbf{l}} \neq \mathbf{l} \vert T_\text{ec})
% \leq p_{\text{ec}}\varepsilon_{\text{cor}}^{\text{cond}}, \label{BIS_ecorrectness}
% \end{equation}
% where $\varepsilon_{\text{cor}}^{\text{cond}}$ now represents the \textit{conditional} correctness (bounding the probability that Alice's and Bob's corrected
% sequences are different, even if their hashes coincide).
% Similarly, we can rewrite Eq.~\eqref{esecrecy} as follows
% \begin{equation}
%    p_{\text{ec}}D(\bar{\rho}^{n}_{BEF|T_\text{ec}},\omega^{n}_{B} \otimes \bar{\rho}^{n}_{EF|T_\text{ec}})\leq  p_{\text{ec}} \varepsilon_{\text{sec}}^{\text{cond}}, \label{BIS_esecrecy}
% \end{equation}
% where $\varepsilon_{\text{sec}}^{\text{cond}}$ bounds the trace distance \textit{conditionally} to successful EC.
% With these refinements, the total epsilon security in Eq.~\eqref{eps_sec_total} is given by 
% \begin{equation}
% \varepsilon=p_\text{ec} \varepsilon_{\mathrm{tot}},~~\varepsilon_{\mathrm{tot}}:= \varepsilon_{\text{cor}}^{\text{cond}}+\varepsilon_{\text{sec}}^{\text{cond}}+ \varepsilon_\text{ent}+n_{\text{pm}}\varepsilon_{\text{pe}},\label{BIS_eps_sec_total}
% \end{equation}
% where all the epsilons in $\varepsilon_{\mathrm{tot}}$ are conditional error probabilities (conditional to successful EC).

For the secret key rate, we may consider the conditional or unconditional secret key rate. 
The conditional rate $R_{\mathrm{cond}}$ is defined on a successfully-corrected block, while the unconditional rate $R$ is the average output rate (the one considered in the previous sections of the paper). Note that we may write
\begin{equation}
R_{\mathrm{cond}} =\frac{s_{n}}{N},~ R=p_{\text{ec}}R_{\mathrm{cond}}, ~R_{\mathrm{ses}}=R.\label{BIS_key_rate_final}
\end{equation}

Finally, another important clarification regards PE. Our theory does not depend on how estimators ($\hat{T}$ and $\hat{\xi}$) and worst-case values ($T_\mathrm{wc}$ and $\xi_\mathrm{wc}$) are calculated. In fact, for any QKD session, these quantities can be sampled and computed directly from the available experimental data. For theoretical investigations, one can adopt the derivations presented in Appendix~\ref{PEtailbounds} but other estimation methods can be employed.

\bigskip

\section{Conclusions\label{SEC4}}
In this paper, we have introduced an improved formulation for the composable and finite-size secret key rate of a generic CV-QKD protocol. By resorting to previous theory and proving various other tools, such as a refined tensor-product reduction for the state after error correction, we have derived simpler and more optimistic formulas, able to show an improvement in the general performance of CV-QKD. As shown in the examples, this improvement can be appreciated both in terms of increased robustness to loss and/or reduced requirements for the size of the usually larger QKD blocks. In general, this work contributes to making a step forward in the rigorous deployment of CV-QKD protocols in practical scenarios.

\section*{Acknowledgements}
This work was supported by the EPSRC via the UK Quantum Communications Hub (Grant No. EP/T001011/1).

%\newpage
\appendix
\section{Proof of the tensor-product reduction in Eq.~\eqref{step_interim} \label{appEXT}}
Consider an arbitrary Hilbert space $\mathcal{H}$ and\ two
generally sub-normalized states $\rho,\tau\in
S_{\leq}(\mathcal{H})$ with
$\mathrm{Tr}\rho,\mathrm{Tr}\tau\leq1$. We may consider the
purified distance~\cite{PurifiedDist}
$P(\rho,\tau)=\sqrt{1-F_{G}(\rho,\tau)^{2}}$, where
$F_{G}$ is the generalized quantum fidelity~\cite[Def. 3.2,
Lemma~3.1]{TomaThesis}.

For any (generally sub-normalized) state $\rho$ of two quantum
systems $A$ and $B$, we may
write~\cite[Def. 5.2]{TomaThesis}%
\begin{equation}
H_{\text{min}}^{\varepsilon}(A|B)_{\rho}=\max_{\tau\in\mathcal{B}%
^{\varepsilon}(\rho)}H_{\text{min}}(A|B)_{\tau}, \label{torrrrGG}%
\end{equation}
where
\begin{equation}
\mathcal{B}^{\varepsilon}(\rho):=\{\rho^{\prime}:\mathrm{Tr}\rho^{\prime}%
\leq1,P(\rho^{\prime},\rho)\leq\varepsilon<1\}
\end{equation}
is a ball of generally sub-normalized states around $\rho$. In
particular, for any generally sub-normalized CQ state 
\begin{equation}
\rho_{CQ}=\sum_{x}p_x|x\rangle_C\langle x|\otimes \rho^x_{Q},
\end{equation}
for $x$ in the alphabet $\mathcal{X}$, we can find another generally sub-normalized CQ state
$\tau_{CQ}\in\mathcal{B}^{\varepsilon}(\rho_{CQ})$ such
that~\cite[Prop.
5.8]{TomaThesis}%
\begin{equation}
H_{\text{min}}^{\varepsilon}(C|Q)_{\rho}=H_{\text{min}}(C|Q)_{\tau}.
\label{torrr}%
\end{equation}
In particular, we may write
\begin{equation}\label{optimum state}
\tau_{CQ}=\sum_{x}q_x|x\rangle_C\langle x|\otimes \tau^x_{Q}.
\end{equation}
Let us now consider a CCQ extension of $\rho_{CQ}$ denoted by $\rho_{C'CQ}$ such as $\rho_{CQ}=\text{Tr}_{C'}(\rho_{C'CQ})$. More specifically, we may write
\begin{align}
&\rho_{C'CQ}=\sum_{x',x} p_{x',x} |x',x\rangle_{C'C}\langle x', x|\otimes \rho_Q^{x',x},\\
&\rho_{CQ}=\sum_{x',x} p_{x',x} |x\rangle_{C}\langle x|\otimes \rho_Q^{x',x},\label{main_state}
\end{align}
where the summation takes place over all the elements $x' \in \mathcal{X}'$ and $x \in \mathcal{X}$ of the basis $\{|x',x\rangle_{C'C}\langle x', x|\}$.

Then there is an extension for $\tau_{CQ}$~\cite[Col.~9]{duality_MT&RR}, denoted by  $\bar{\tau}_{C'CQ}$, such that
\begin{equation}\label{eq_pd_first}
P(\bar{\tau}_{C'CQ},\rho_{C'CQ})=P(\tau_{CQ},\rho_{CQ}) \leq \varepsilon. 
\end{equation}
Note that, due to the monotonicity of the purified distance~\cite[Theorem~3.4]{TomaThesis} under CPTP maps $\mathcal{E}$, we have
\begin{align}\label{ineq_pd_sec}
P\left( \mathcal{E}(\bar{\tau}_{C'CQ}),\mathcal{E}(\rho_{C'CQ})\right)\leq P(\bar{\tau}_{C'CQ},\rho_{C'CQ}).
\end{align}

In particular, consider a ``pinching'' channel~\cite[Def.~4.4]{JWQIT} in the basis $\{|x',x\rangle_{C'C}\langle x', x|\}$, i.e.,
\begin{equation}
\mathcal{E}_\text{pch}(\rho)=\sum_{x',x}|x',x\rangle_{C'C}\langle x', x|\rho|x',x\rangle_{C'C}\langle x', x|.
\end{equation}
This channel transforms an arbitrary input state into an output CCQ state, which is classical in the systems $C'C$, i.e., with respect to the basis  $\{|x',x\rangle_{C'C}\langle x', x|\}$. At the same time, it is clear that this channel does not change $\rho_{C'CQ}$. According to Sec.~\ref{CCQ_FORM}, we may write 
\begin{align}
\tau_{C'CQ}&=\mathcal{E}_\text{pch}(\bar{\tau}_{C'CQ})\notag\\
&=\sum_{x',x} q_{x',x} |x',x\rangle_{C'C}\langle x', x|\otimes \tau_Q^{x',x},\label{CCQ_EXP}\\
\tau_{CQ}&=\sum_{x',x} q_{x',x} |x\rangle_{C}\langle  x|\otimes \tau_Q^{x',x},\label{after_expansion}
\end{align}
% where $\Pi_{C'}=\sum_{x'}|x'\rangle_{C'}\langle x'|$ for $\{|x'\rangle\langle x|\}$ being the computational basis for $C'$. 
and we have
\begin{equation}
P\left( \tau_{C'CQ},\rho_{C'CQ}\right)=P\left( \mathcal{E}_\text{pch}(\bar{\tau}_{C'CQ}),\mathcal{E}_\text{pch}(\rho_{C'CQ})\right)\leq \varepsilon,
\end{equation}
as a consequence of Eqs.~\eqref{eq_pd_first} and~\eqref{ineq_pd_sec} specified to the pinching channel.
% This means that any extension with no classical structure for $C'$ will have a smaller purified distance with $\bar{\rho}$, which remains the same after the application of $\Pi_{C'}$. Therefore, we may write
% \begin{align}
% &\bar{\tau}_{C'CQ}=\sum_{x',x} q_{x',x} |x',x\rangle_{C'C}\langle x',x|\otimes \tau_Q^{x',x}\\
% &\tau_{CQ}=\sum_{x',x} q_{x',x} |x\rangle_{C}\langle x|\otimes \tau_Q^{x',x}\label{tau_state}
% \end{align}

Consider the joint projection 
\begin{equation}
    \Pi:=\sum_{(x', x)\in\Gamma}\left\vert
x',x\right\rangle _{C'C}\left\langle x',x\right\vert,
\end{equation}
defined over a
reduced alphabet $\Gamma \subseteq \mathcal{X'}\otimes \mathcal{X}$ for the classical system
$C'C$ and a subsequent guess channel $\cal{E}_{\mathrm{guess}}$ applied to $C'C$ [cf. Eqs.~\eqref{EC-proj} and~\eqref{chanMain} in the main text]. Then due to the monotonicity of the purified distance under completely positive trace non-increasing maps, i.e., projections, CPTP maps, and partial trace operations~\cite[Theorem~3.4]{TomaThesis}, we have 
\begin{align}
&P(\tilde{\tau}_{CQ},\tilde{\rho}_{CQ})\leq P\left( \tau_{C'CQ},\rho_{C'CQ}\right)\leq \varepsilon,
\end{align}
where
\begin{align}
\tilde{\rho}_{CQ}&=\text{Tr}_{C'} [ \mathcal{E}_{\text{guess}}  \left( \Pi \rho_{C'CQ}\Pi \right) ]\notag\\
&=\sum_{(x',x) \in \Gamma} p_{x',x} |x\rangle_{C}\langle x|\otimes \rho_Q^{x',x}\label{tilde_state},\\
\tilde{\tau}_{CQ}&=\text{Tr}_{C'} [ \mathcal{E}_{\text{guess}} \left(\Pi\tau_{C'CQ}\Pi\right) ]\notag\\
&=\sum_{(x',x) \in \Gamma} q_{x',x} |x\rangle_{C}\langle x|\otimes \tau_Q^{x',x}.\label{tildetau_state}
\end{align}
This means that $\tilde{\tau}_{CQ} \in \mathcal{B}^\varepsilon(\tilde{\rho}_{CQ})$ and as a consequence of the definition of the smooth min-entropy
\begin{equation}\label{tilde_entropy_rel}
H_\text{min}^\varepsilon(C|Q)_{\tilde{\rho}} \geq H_\text{min}(C|Q)_{\tilde{\tau}}.
\end{equation}

Then we exploit the following formula~\cite[Sec.~4.2.1]{TomaThesis} for the min-entropy
\begin{equation}
H_\text{min}(A|B)_\tau=-\log_2 \underset{\mathcal{E}}{\operatorname{max}}~\mathrm{Tr}\left[\mathcal{E}_{B\rightarrow B'}(\tau_{AB}) \gamma_{AB'}\right],
\end{equation}
where $\tau_{AB}\in \mathcal{S}_{\leq}(\mathcal{H}_{AB})$ is a sub-normalized state for systems $A$ and $B$, $\gamma_{AB'}=|\gamma_{AB'}\rangle\langle \gamma_{AB'}|$ is a sub-normalized maximally-entangled state for systems $A$ and $B'$, i.e.,
\begin{equation}
|\gamma_{AB'}\rangle=\sum_x |x\rangle \otimes |x\rangle,
\end{equation}
and $\mathcal{E}_{B\rightarrow B'}$ is a CPTP map from $B$ to $B'$, where $\mathcal{H}_{B'}\cong\mathcal{H}_{A}$. %with $d$ being the dimension of the Hilbert space. 
In particular, when we assume CQ states as in Eq.~\eqref{after_expansion}, we may write
\begin{align}
&\text{Tr}\left[\mathcal{E}_{Q\rightarrow Q'}(\tau_{CQ}) \gamma_{CQ'} \right]\notag\\
&=\sum_{x',x}q_{x',x}\text{Tr}\left[|x\rangle_{C}\langle x|\otimes \mathcal{E}_{Q\rightarrow Q'}(\tau_Q^{x',x})\gamma_{CQ'}\right]\notag\\
&=\sum_{x',x}q_{x',x} \langle x  | \mathcal{E}_{Q\rightarrow Q'}(\tau_Q^{x',x}) | x \rangle \notag \\
&\geq \sum_{(x',x) \in \Gamma}q_{x',x} \langle x  | \mathcal{E}_{Q\rightarrow Q'}(\tau_Q^{x',x}) | x \rangle \notag \\
%&\geq \sum_{(x',x) \in \Gamma}q_{x',x}\text{Tr}\left[|x\rangle_{C}\langle x|\otimes \mathcal{E}_{Q\rightarrow Q'}(\tau_Q^{x',x})\Gamma_{CQ'}\right]\notag\\
&=\text{Tr}\left[\mathcal{E}_{Q\rightarrow Q'}(\tilde{\tau}_{CQ})\gamma_{CQ'}\right],
\end{align}
where the inequality stems from the fact that we have a summation of a smaller amount of positive terms due to the reduced alphabet $(x',x) \in \Gamma$ of the projection.
By taking the maximum and the minus logarithm of the previous relation, we may write the following relation for the min-entropies of the states $\tau_{CQ}$ and $\tilde{\tau}_{CQ}$:
\begin{equation}
H_\text{min}(C|Q)_{\tilde{\tau}}\geq H_\text{min}(C|Q)_{\tau}.
\end{equation}

By replacing Eq.~\eqref{torrr} and~\eqref{tilde_entropy_rel} in the previous inequality, we obtain the corresponding inequality for the smooth min-entropies of $\rho$ and $\tilde{\rho}$, i.e.,
\begin{equation}
H^\varepsilon_\text{min}(C|Q)_{\tilde{\rho}}\geq H^\varepsilon_\text{min}(C|Q)_{\rho}.
\end{equation}

Finally, we note that we get Eq.~\eqref{step_interim}, by
replacing $\rho \rightarrow \rho^{\otimes n}$, $\tilde{\rho} \rightarrow
\sigma^{n}$, $C \rightarrow B^n$, and $Q \rightarrow E^n$.

\subsection{Form of the CCQ extension in Eq.~\eqref{CCQ_EXP}\label{CCQ_FORM}}
Let us assume a general CCQ state 
\begin{align}\label{theta}
\theta_{C'CQ}&=\sum_{x',x} \tilde{q}_{x}\tilde{q}_{x'|x} 
|x',x\rangle_{C'C} \langle x',x|\otimes \tilde{\tau}_{Q}^{x',x}
\end{align}
and we set
\begin{equation}
\tilde{\tau}_{Q}^x:=\sum_{x} \tilde{q}_{x'|x}\tilde{\tau}^{x',x}_{Q}.
\end{equation}
We impose that the reduced state, after tracing out $A$, is equal to Eq.~\eqref{optimum state}. From the block diagonal form of the states, we obtain
\begin{equation}\label{state_eq}
q_x \tau^x_{Q}=\tilde{q}_x\tilde{\tau}_{Q}^x.
\end{equation}
Similarly, by further tracing out $E$, we have that
\begin{equation}\label{prob_eq}
q_x\text{tr}\{\tau_{Q}^x\}=\tilde{q}_x\text{tr}\{\tilde{\tau}_{Q}^x\}.
\end{equation}
By combining Eqs.~\eqref{state_eq} and~\eqref{prob_eq}, we obtain
\begin{align}
\tau^x_{Q}&=\frac{\text{tr}\{\tau_{Q}^x \}}{\text{tr}\{\tilde{\tau}_{Q}^x \}}\tilde{\tau}^x_{Q}\notag\\
&=\sum_{x'}\tilde{q}_{x'|x}\frac{\text{tr}\{\tau_{Q}^x \}}{\text{tr}\{\tilde{\tau}_{Q}^x \}}\tilde{\tau}_{Q}^{x',x}.\label{tauQx}
% \\
% &=\sum_{x' }q_{x'|x}\tau_{Q}^{x',x}
\end{align}
We can freely set 
\begin{align}
q_{x'|x}&:=\frac{\tilde{q}_{x'|x}}{\text{tr}\{\tilde{\tau}_{Q}^x \}}\label{q_rep},\\
\tau_{Q}^{x',x}&:=\text{tr}\{\tau_{Q}^x \}\tilde{\tau}_{Q}^{x',x}\label{tau_rep},
\end{align}
so Eq.~\eqref{tauQx} simply becomes
\begin{equation}
\tau^x_{Q}=\sum_{x' }q_{x'|x}\tau_{Q}^{x',x}.
\end{equation}
Then, by using Eq.~\eqref{prob_eq} in Eq.~\eqref{theta}, we obtain
\begin{align}\label{tau}
\theta_{C'CQ}=\sum_{x',x} &q_{x} \text{tr}\{\tilde{\tau}_{Q}^x\}^{-1}
 \tilde{q}_{x'|x}|x',x\rangle_{C'CQ} \langle x',x|\notag\\
 &\otimes \text{tr}\{\tau_{Q}^x\}\tilde{\tau}_{Q}^{x',x}\notag\\
 =\sum_{x',x} &q_{x}q_{x'|x} |x',x \rangle_{C'CQ} \langle x', x| \otimes \tau_{Q}^{x',x},
\end{align}
where, in the last equation, we have used Eqs.~\eqref{q_rep} and~\eqref{tau_rep}. Note that the fact that the state must be CCQ and that its reduced form must be equal to Eq.~\eqref{optimum state} completely characterizes the state. Therefore, we can derive the form in Eq.~\eqref{CCQ_EXP}.

\section{Details on parameter estimation for Gaussian-modulated protocols \label{PEtailbounds}}
We assume that $V_0 m$ data points are used for PE, with $V_0=1$ ($V_0=2$) for the homodyne (heterodyne) protocol. For simplicity, we assume that the two quadratures have been modulated with the same variance and that the channel transforms them in the same way (phase-insensitive channel, as typical of the standard thermal-loss channel). %This means that, in the heterodyne protocol, we have a double number of variables.  
Then we denote with $x$ and $y$ the Gaussian input and output of the channel, respectively,  with Gaussian noise variable $z$ and transmissivity $T$, where
\begin{equation}
y=\sqrt{\eta T}x+z.
\end{equation}
\SP{Note that, in this appendix, we adopt a different notation for the heterodyne protocol. In the main text, $y$ represented both Bob's quadratures, i.e., $y=(q_B,p_B)$. Here, $y$ represents Bob's generic quadrature, i.e. $y=q_B$ or $p_B$.}

% Note that, in this appendix, we adopt a different notation. In the main text, $x$ and $y$ represented Alice's and Bob's quadrature variables used in the protocol. For example, for heterodyne, this means $y=(q_B,p_B)$. Here, $x$ and $y$ represent Alice's and Bob's generic quadrature variables. For heterodyne, this means $y=q_B$ or $p_B$.

\subsection{Estimating the transmissivity}
We write the covariance $C_{xy}=\text{Cov}(x,y)=\sqrt{\eta T}\sigma_x^2$, where $\sigma_x^2$ is the variance of $x$. Its estimator is given by 
\begin{align}
 \widehat{C}_{xy}:&=\frac{1}{V_0 m}\sum_{i=1}^{V_0m}[x]_i[y]_i\label{strict_oneway}\\&=\frac{1}{V_0 m}\sum_{i=1}^{V_0 m}\sqrt{T}[x]_i^2+[x]_i[z]_i\notag\\&\simeq\sqrt{\eta T}\sigma^2_x+\frac{1}{V_0 m}\sum_{i=1}^{V_0m}[x]_i[z]_i,\label{simple_oneway}
\end{align}
where, in Eq.~\eqref{simple_oneway}, we replaced Alice's known variance. 
%In this case, it does not contribute to the variance of $\widehat{C}_{xy}$. 

We calculate $V_\text{Cov}:=\text{Var}(\widehat{C}_{xy})$ directly from Eq.~\eqref{strict_oneway} and obtain
\begin{align}
V_\text{Cov}&=\frac{1}{V_0 m}\left[\eta T\text{Var}(x^2)+\sigma_x^2\sigma_z^2\right]\notag\\
&=\frac{1}{V_0 m}\left[\eta T2(\sigma_x^2)^2+\sigma_x^2\sigma_z^2\right]\notag\\
&=\frac{1}{V_0 m}\eta T(\sigma_x^2)^2\left[2+\frac{\sigma_z^2}{ \eta T\sigma_x^2}\right].
\end{align}
Otherwise, we can start from Eq.~\eqref{simple_oneway} and obtain
\begin{equation}
V_\text{Cov}=\frac{1}{V_0 m}\sigma^2_x\sigma_z^2.
\end{equation}
Both of them can be summarized into
\begin{equation}
V_\text{Cov}=\frac{C^2_{xy}}{V_0 m}\left[c_\text{pe}+\frac{\sigma_z^2}{\eta T\sigma_x^2}\right],
\end{equation}
where, for the first one, we set $c_\text{pe}=2$ and, for the second one, $c_\text{pe}=0$.
Then we may write the estimator
\begin{equation}\label{Tau_est}
\widehat{T}=\frac{1}{\eta(\sigma_x^2)^2}\hat{C}^2_{xy}=\frac{V_\text{Cov}}{\eta(\sigma_x^2)^2}\left(\frac{\hat{C}_{xy}}{\sqrt{V_\text{Cov}}}\right)^2.
\end{equation}

For the central limit theorem (CLT), the quantity $\frac{\widehat{C}_{x,y}}{\sqrt{V_\text{Cov}}}$ tends to a Gaussian. In fact, we can bound the Kolmogorov distance of its actual distribution from a Gaussian distribution. For typically-large block sizes, one can use the Berry-Esseen inequality to check, numerically, that the distance becomes quickly negligible.

Since $\frac{\widehat{C}_{x,y}}{\sqrt{V_\text{Cov}}}$ follows a standard normal distribution with mean $\frac{C_{x,y}}{\sqrt{V_\text{Cov}}}$, then $\left(\frac{\widehat{C}_{x,y}}{\sqrt{V_\text{Cov}}}\right)^2$  follows a non-central chi-squared distribution with degrees of freedom $d_f=1$ and non-centrality parameter $\kappa_{cn}=C_{x,y}^2/V_\text{Cov}$. Consequently $\widehat{T}$ follows the same distribution but rescaled by the factor $\frac{V_\text{Cov}}{\eta(\sigma_x^2)^2}$. Via the chi-squared distribution parameters, we can calculate its variance 
\begin{equation}
\text{Var}(\widehat{T})=\frac{2 V_\text{Cov}^2}{\eta^2(\sigma_x^2)^4}\left(1+2\frac{C_{x,y}^2}{V_\text{Cov}}\right),
\end{equation}
and, by omitting the terms $\mathcal{O}(1/m^2)$, we obtain
\begin{align}\label{varThat}
\text{Var}(\widehat{T})&=\frac{4V_\text{Cov}C^2_{x,y}}{\eta^2(\sigma_x^2)^4}=\frac{4C^4_{x,y}}{\eta^2(\sigma_x^2)^4}\frac{\left[c_\text{pe}+\frac{\sigma_z^2}{\eta T\sigma_x^2}\right]}{V_0 m}\\=&\frac{4\eta^2T^2(\sigma_x^2)^4}{\eta^2
(\sigma_x^2)^4}\frac{\left[c_\text{pe}+\frac{\sigma_z^2}{\eta T\sigma_x^2}\right]}{V_0m}\notag\\
&=\frac{4T^2}{V_0m}\left[c_\text{pe}+\frac{\sigma_z^2}{\eta T\sigma_x^2}\right]:=\sigma_T^2.
\end{align}
Given that
\begin{equation}
    \sigma_z^2=\eta T \xi+u_\text{el}+V_0,\label{excessConnection}
\end{equation}
we may write
\begin{equation}
    \sigma_T=\frac{2T}{\sqrt{V_0m}}\sqrt{\left[c_\text{pe}+\frac{\xi+\frac{V_0+u_\text{el}}{\eta T}}{\sigma_x^2}\right]},
\end{equation}
as in Eq.~\eqref{sigma_tau} of the main text up to replacing \SP{$V=\sigma^2_x$}. 

For large enough $m\gg1$, we have again a good convergence in the CLT and, therefore, we may assume that the distribution of $\widehat{T}$ becomes Gaussian with variance given by Eq.~\eqref{varThat}. Therefore, we may write that
\begin{equation}\label{Twc_generic}
T_\text{wc} \simeq T-w \sigma_T
\end{equation}
with 
\begin{equation}
w=\sqrt{2}\text{erf}^{-1}(1-2\varepsilon_\text{pe}). \label{PE-w}
\end{equation}

To better understand the result above, let us assume a generic estimator $\widehat{p}$ that follows a normal distribution with mean $p$ and variance $\sigma_p^2$. We impose the probability that $\widehat{p}\geq p_\text{wc}:=p +w \sigma_p$
is less than $\varepsilon_\text{pe}$. In other words,  
\begin{equation}\label{prob_of_error}
\mathrm{Prob}[\widehat{p}\geq p +w \sigma_T]\leq \varepsilon_\text{pe}.
\end{equation}
% \begin{remark}
% The case for $\widehat{p}\leq p_\text{wc}:=p -w \sigma_p$ is equivalent to the previous bound due to the symmetry of the Gaussian distribution.
% \end{remark}
% Then we bring the previous relation into the corresponding standard distribution form by applying the following calculations
We can re-write Eq.~\eqref{prob_of_error} as follows
\begin{align}
\mathrm{Prob}\left[\widehat{p}-p\geq w \sigma_p\right]\leq \varepsilon_\text{pe}\notag\\
\mathrm{Prob}\left[\frac{\widehat{p}-p}{\sigma_p}\geq w \right]\leq \varepsilon_\text{pe}.
\end{align}
We can recognize the cumulative distribution 
\begin{equation}
\Phi(w)=\frac{1}{2}\left[1+\text{erf}(w/\sqrt{2})\right]
\end{equation} 
of the normal variable $\frac{\widehat{p}-p}{\sigma_p}$. We use its connection to the error function $\text{erf}(.)$ to write
\begin{align}
1-\Phi(w)\leq \varepsilon_\text{pe},\\
%-\Phi(w)\leq \varepsilon-1\\
%\Phi(w)\geq 1-\varepsilon\\
\frac{1}{2}\left[1+\text{erf}(w/\sqrt{2})\right]\geq 1-\varepsilon_\text{pe},\\
%1+\text{erf}(w/\sqrt{2})\geq 2 (1-\varepsilon)\\
%\text{erf}(w/\sqrt{2})\geq 2-2\varepsilon-1\\
\text{erf}(w/\sqrt{2})\geq 1-2\varepsilon_\text{pe},\\
%\frac{w}{\sqrt{2}}\geq \text{erf}^{-1}(1-2\varepsilon_\text{pe})\\
w\geq \sqrt{2} \text{erf}^{-1}(1-2\varepsilon_\text{pe}),
\end{align}
amd we use the bound above in Eq.~\eqref{PE-w}.

Alternatively, we may use tail bounds for the chi-squared distribution. In particular, for the stochastic variable $X$ following the latter distribution, we have that 
\begin{align}\label{tail_bounds}
&\text{Prob} \left [ X \leq (d_f+\kappa_\text{nc}) -2\sqrt{(d_f+2\kappa_\text{nc})\ln \varepsilon^{-1}_\text{pe}} \right]  \leq \varepsilon_\text{pe},\\
&\text{Prob} \Big [ X \geq (d_f+\kappa_\text{nc}) +2\sqrt{(d_f+2\kappa_\text{nc})\ln \varepsilon^{-1}_\text{pe}}\notag\\&~~~~~~~~~~~~~~~~~~~~~~~~~~~~~~~~~~~~~~~~+ 2\ln \varepsilon_\text{pe}^{-1}\Big]  \leq \varepsilon_\text{pe}.
\end{align}
Applying this to $\widehat{T}$, we obtain
\begin{align}
T_\text{wc} &= \frac{V_\text{Cov}+C_{x,y}^2}{\eta(\sigma_x^2)^2}\notag\\&~~~ -\frac{2}{\eta(\sigma_x^2)^2}\sqrt{(V_\text{Cov}^2+2C_{x,y}^2V_\text{Cov})\ln \varepsilon^{-1}_\text{pe}}.
\end{align}
Then, we expand the square root above and omit $\mathcal{O}(\frac{1}{m})$ terms 
\begin{align}
&\sqrt{(V_\text{Cov}^2+2C_{x,y}^2V_\text{Cov})}=\sqrt{2C_{x,y}^2V_\text{Cov}}\sqrt{1+\frac{V_\text{Cov}}{2C_{x,y}^2}}\notag\\
&=\sqrt{2C_{x,y}^2V_\text{Cov}}\left[\left(1+\frac{V_\text{Cov}}{4C_{x,y}^2}\right)+\mathcal{O}\left(\frac{1}{m^2}\right)\right]\notag\\
&=\sqrt{2C_{x,y}^2V_\text{Cov}}+\mathcal{O}\left(\frac{1}{m}\right)\notag\\
&\simeq\sqrt{2\frac{\eta^2T^2(\sigma_x^2)^4}{V_0m}\left[c_\text{pe}+\frac{\sigma_z^2}{\eta T\sigma_x^2}\right]}.
\end{align}
Finally, we obtain
\begin{align}
&T_\text{wc} \simeq T -\sqrt{2\ln \varepsilon^{-1}_\text{pe}}\frac{2T}{\sqrt{V_0m}}\sqrt{\left[c_\text{pe}+\frac{\sigma_z^2}{\eta T\sigma_x^2}\right]},
\end{align}
which can be written in the form of Eq.~\eqref{Twc_generic} but with 
\begin{equation}
w=\sqrt{2\ln \varepsilon^{-1}_\text{pe}}.\label{otherW}
\end{equation}

\subsection{Estimating the noise\label{est_noise}}

In the same manner, we calculate the estimator for $\sigma_z^2$, the variance of the noise variable $z$. We have that
\begin{align}
\widehat{\sigma}_z^2=&\frac{1}{V_0m}\sum_{i=1}^{V_0m}\left(y-\sqrt{\eta\widehat{ T}}x\right)^2\label{noise_est}\\
\simeq&\frac{1}{V_0m}\sigma_z^2\sum_{i=1}^{V_0m}\left(\frac{y-\sqrt{\eta T}x}{\sigma_z}\right)^2.
\end{align}
The sum above follows a central chi-squared distribution with $d_f=V_0m$ and, therefore, with mean $V_0m$ and variance $2V_0m$. Then $\widehat{\sigma}_z^2$ follows the same distribution but rescaled by $\frac{\sigma_z^2}{V_0m}$. Thus, its mean value is $\sigma_z^2$ while its variance $ V_z= \frac{2(\sigma_z^2)^2}{V_0m}$. From this, we may write
\begin{equation}\label{sigmaz_gaussian}
[\sigma_z^2]_\text{wc} \simeq \sigma_z^2+w\sqrt{V_z}
\end{equation}
with $w$ given by Eq.~\eqref{PE-w}. 

Otherwise, we may use the tail bounds in Eq.~\eqref{tail_bounds} to obtain
\begin{align}
[\sigma_z^2]_\text{wc}&=\frac{\sigma_z^2}{V_0m}\left(V_0m+2\sqrt{V_0m\ln\varepsilon_\text{pe}^{-1}}+2\ln\varepsilon_\text{pe}^{-1}\right)\notag\\
=&\sigma_z^2+\sigma_z^2\frac{\sqrt{2}}{\sqrt{V_0m}}\sqrt{2\ln\varepsilon_\text{pe}^{-1}}+\mathcal{O}\left(\frac{1}{m}\right)\notag\\
\simeq&\sigma_z^2+\sigma_z^2\frac{\sqrt{2}}{\sqrt{V_0m}}\sqrt{2\ln\varepsilon_\text{pe}^{-1}},
\end{align}
which can be written as in Eq.~\eqref{sigmaz_gaussian} but with $w$ given in Eq.~\eqref{otherW}.%$$w=\sqrt{2\ln\varepsilon_\text{pe}^{-1}}.$

Finally, from Eq.~\eqref{excessConnection}, we derive
\begin{align}
\xi_\text{wc}=&\frac{[\sigma_z^2]_\text{wc}}{\eta T_\text{wc}}-\frac{u_\text{el}+V_0}{\eta T_\text{wc}}\notag\\
&\simeq\frac{\eta T \xi+ w\sqrt{V_z}+u_\text{el}+V_0}{\eta T_\text{wc}}-\frac{u_\text{el}+V_0}{\eta T_\text{wc}}\notag\\
&=\frac{T}{T_\text{wc}}\xi+\frac{w\sqrt{V_z}}{\eta T_\text{wc}}.
\end{align}
This expression can equivalently be written as
\begin{equation}\label{generic_xi_wc}
    \xi_\text{wc} \simeq \frac{T}{T_\text{wc}}\xi+w \sigma_{\xi},
\end{equation}
where
\begin{equation}
     \sigma_{\xi}=\frac{\sqrt{V_z}}{\eta T_\text{wc}}=\sqrt{\frac{2}{V_0m}} \frac{\eta T \xi+V_0+u_\text{el}}{\eta T_\text{wc}},
\end{equation}
as in Eq.~\eqref{sigma_xi} of the main text.
\newpage

\begin{thebibliography}{99}
\bibitem{endnote2} Note that, for the heterodyne protocol, the estimation of Bob's entropy can also be done on the variables before concatenation as in Refs.~\cite{Alex_II,Alex_III}.
\end{thebibliography}
\end{document}